\documentclass[twocolumn,tighten]{aastex6}
\turnoffedit

\usepackage[preserveurlmacro]{breakurl}

\begin{document}

\title{Cosmic Infrared Background Fluctuations and Zodiacal Light}
\author{Richard G. Arendt\altaffilmark{1}, A. Kashlinsky\altaffilmark{2},
S. H. Moseley\altaffilmark{3}, J. Mather\altaffilmark{3}}
\affil{Observational Cosmology Laboratory, Code 665,
Goddard Space Flight Center, 8800 Greenbelt Road, Greenbelt, MD
20771, USA; Richard.G.Arendt@nasa.gov, 
Alexander.Kashlinsky@nasa.gov,
Harvey.Moseley@nasa.gov, John.C.Mather@nasa.gov}


\begin{abstract}
We have performed a specific observational test to measure the 
effect that the zodiacal light can have on measurements of the spatial 
fluctuations of the near-IR background. Previous estimates of possible
fluctuations caused by zodiacal light have often been extrapolated from 
observations of the thermal emission at longer wavelengths and 
low angular resolution, or from IRAC observations of high latitude 
fields where zodiacal light is faint and not strongly varying with time.
The new observations analyzed here target the COSMOS field, at low
ecliptic latitude where the zodiacal light intensity varies by factors 
of $\sim2$ over the range of solar elongations at which the field can 
be observed. We find that the white noise component of the spatial 
power spectrum of the background is correlated with the 
modeled zodiacal light intensity. \edit1{Roughly half of the 
measured white noise is correlated with the zodiacal light, but 
a more detailed interpretation of the white noise is hampered
by systematic uncertainties that are evident in the zodiacal light model.} 
At large angular scales ($\gtrsim100''$) where excess power 
above the white noise is observed, we find no correlation of the 
power with the modeled intensity of the zodiacal light. This test 
clearly indicates that the large scale power in the infrared background 
is not being caused by the zodiacal light.
\end{abstract}

\keywords{cosmology: observations --- diffuse radiation --- zodiacal dust}

\section{Introduction}

\footnotetext[1]{CRESST/University of Maryland -- Baltimore County}
\footnotetext[2]{Science Systems \& Applications Inc.}
\footnotetext[3]{NASA}
\setcounter{footnote}{3}

The study of astronomical backgrounds at various wavelengths 
allows the examination of sources that are intrinsically diffuse, or 
individually too faint or too confused to be detected. Over time,
improvements in instrumentation may resolve increasingly fainter sources,
but very faint and intrinsically diffuse sources always remain in the realm
of background studies. 

Studies of the cosmic infrared background (CIB) have aimed at 
measuring the cumulative stellar emission of galaxies across \edit1{the} 
entire history of the universe. Measurements made by the DIRBE 
instrument on {\it COBE} provided the first space-based
measurements of the absolute sky surface brightness at wavelengths 
from 1.25 to 240 $\mu$m with an angular resolution of $0\fdg7$ 
\citep{hauser:1998}. However, difficulties in accurately removing 
foreground contributions from the zodiacal light \citep{kelsall:1998}, 
and from Galactic stars and interstellar dust \citep{arendt:1998}
prevented precise detections of the CIB except at the longest wavelengths.

Subsequent measurements and studies of the CIB at near- to mid-IR wavelengths
have made use of telescopes and instruments that provide much higher 
angular resolution than DIRBE, but which usually cover only a small 
fraction of the sky,
e.g. 2MASS, {\it Spitzer}, {\it HST}, {\it IRTS}, {\it AKARI}, {\it CIBER}.
The higher angular resolution allows the exclusion or subtraction of 
stars and bright galaxies from sky brightness measurements, such 
that the contribution of Galactic stars is minimized, and the CIB that 
remains does not include the contributions of galaxies brighter than 
certain limits dependent on depth of the observations.
These data sets are often better suited for measuring the spatial 
fluctuations or structure of the ``source-subtracted'' CIB, 
rather than for measuring the mean value of the CIB
\citep{
kashlinsky:1996,
kashlinsky:1996a,
kashlinsky:2002,
kashlinsky:2005,
kashlinsky:2007b,
kashlinsky:2012,
kashlinsky:2000,
odenwald:2003,
matsumoto:2005,
matsumoto:2015,
matsumoto:2011,
thompson:2007a,
thompson:2007b,
cooray:2007,
sullivan:2007,
arendt:2010a,
matsuura:2011,
pyo:2012,
zemcov:2014,
donnerstein:2015,
mitchell-wynne:2015,
seo:2015}.
These studies seem to show reasonable consistency of 
fluctuation measurements reported by different experiments
and different groups at $\lambda \gtrsim2$ $\mu$m, 
though the picture is less clear at shorter wavelengths.

The \edit1{origin} of the fluctuations is not yet clear.
The fluctuations may arise from any or all of: solar system and 
Galactic foregrounds, nearby extragalactic contributions, and distant 
extragalactic contributions.
In most cases \citep[e.g.][]{kashlinsky:2005,arendt:2010a,matsumoto:2011,zemcov:2014} foregrounds are 
estimated by extrapolation 
of measurements at other wavelengths and locations. This is particularly
true of zodiacal light contributions. Mid-IR observations using {\it ISO} have 
limited zodiacal light fluctuations to $<0.2\%$ on scales $>3'$ \citep{abraham:1997}.
More recent {\it AKARI} measurement set the limit even lower at $0.03\%$
\citep{pyo:2012}. 

Direct detection of zodiacal light influences 
in {\it Spitzer} IRAC measurements
have been checked in existing deep data sets by constructing 
$A-B$ difference maps, where $A$ and $B$ represent observations made at two 
different epochs, typically 6 months and/or 1 year apart 
\citep[e.g.][]{kashlinsky:2007b}. 
The expectation is that the first-order gradient 
\edit1{(and any instrumental imprint)} should reverse at 
6-month intervals, while any smaller scale, physically distinct structures
in the interplanetary dust cloud should not remain fixed in a given field (due 
to differential rotation of the cloud) and should not appear the same
at different epochs separated by 6 months. Excess differences
have not been seen in existing deep CIB studies. However, these 
are generally high latitude fields where the zodiacal light is faint
and not strongly modulated. Additionally, at intervals of 6 months 
and (especially) 1 year, the interplanetary dust cloud may 
be sufficiently symmetric such that certain structures (e.g. those 
associated with the earth-resonant ring) may still cancel out in 
$A-B$ difference maps.

As a more certain test for zodiacal light influences, our new 
{\it Spitzer} IRAC observations
have monitored a low latitude field over an entire visibility window
to obtain a data set where the zodiacal light is both brightest and most 
strongly modulated. The observations were planned to be sufficiently deep 
to detect the reported large scale background structure. Thus 
these data are uniquely suitable for checking if the zodiacal light 
intensity has any effect on the reported background fluctuations at large 
angular scales.

This paper reports on this test and the results. The observations and data
reduction are described in sections \ref{sec:obs} and \ref{sec:data}. Section \ref{sec:power} provides the 
characterization of the power spectra of the background. Section \ref{sec:discuss} discusses
the correlations between the various components of the power spectrum and 
the zodiacal light intensity. Section \ref{sec:summary} summarizes the results. An appendix
provides additional detail on temporal variations in the data that are 
tracked and corrected by the self-calibration procedure that we apply.
The appendix also features additional details on the effects of 
the source model, and on the comparison with previous CIB measurements.

\section{Observations} \label{sec:obs}

Most commonly observed extragalactic fields are chosen to be at high 
Galactic and ecliptic latitudes to minimize the influence 
of foregrounds. \edit1{The observations presented here were
designed, proposed, and approved specifically for the 
purpose of examining the effect of zodiacal light on CIB measurements.}
The COSMOS field \citep{scoville:2007,ashby:2013,ashby:2015}
is suitable for our experiment, as it lies at relatively low 
ecliptic latitude. We have selected a subregion at ecliptic
coordinates $(\lambda,\beta) = (151.73, -8.63)$ which 
is relatively free of bright sources. 
The patch is observed 5 times 
across an entire visibility window of the {\it Spitzer}
spacecraft \citep{werner:2004a,gehrz:2007}, 
covering the widest possible range of solar elongation,
and thus brightness. The size, $\sim10'\times 10'$, and depth, 
$\sim4$~hr per epoch, of the patch were chosen to be 
sufficiently wide and deep to distinguish the 
large scale fluctuations above the random 
white noise in the observations.
We use {\it Spitzer's} IRAC instrument 
\citep{fazio:2004} to collect 3.6 and 4.5 $\mu$m data.
As IRAC's field of view is $5' \times 5'$, the observations
require mosaicking $2 \times 2$ fields of view and a total
observing time of $\sim16$~hr at each epoch.
Table \ref{tab:obs} lists the dates, solar elongations, and astronomical 
observation request (AOR) numbers of the observations.

\begin{deluxetable}{lcccD}
\tabletypesize{\scriptsize}
\tablewidth{0pt}
\tablecaption{COSMOS Zodiacal Light Observations\label{tab:obs}}
\tablehead{
\colhead{Epoch} &
\colhead{Date} &
\colhead{MJD} &
\colhead{AOR} & 
\twocolhead{Solar Elongation}
}
\decimals
\startdata
1 & 2013 Jan 26 & 56318.111 & 42306048 &  \phn\phn\phn\phn83.6 \\
2 & 2013 Feb 02 & 56325.130 & 42306304 &  90.3 \\
3 & 2013 Feb 13 & 56336.289 & 42306560 & 101.0 \\
4 & 2013 Feb 24 & 56347.142 & 42306816 & 111.5 \\
5 & 2013 Mar 02 & 56353.211 & 42307072 & 117.3 \\
\enddata
\tablecomments{MJD = modified julian date, AOR = 
astronomical observation request; Program ID = 80062}
\end{deluxetable}
 
\section{Data Reduction} \label{sec:data}
\subsection{Self-Calibration}
We reduce the data using the same self-calibration 
techniques \citep{fixsen:2000} that have been previously 
employed \citep{arendt:2010a}. The data reduction began with
the IRAC cBCD individual frames and applied a data model of 
\begin{equation}\label{eq:selfcal}
D^i = S^\alpha + F^p +F^q
\end{equation}
where $D^i$ is the measured intensity at pixel $i$ in a single frame ($q$),
$S^\alpha$ is the true sky intensity at location $\alpha$, $F^p$ is an 
offset for pixel $i$ which is constant over all frames, and $F^q$ is an offset
for frame $q$ which is constant over all pixels (but variable with time).
Each of the 5 epochs is self-calibrated separately to provide detector 
offsets, $F^p$, that are appropriate for each epoch. Sky maps are generated on 
a pixel scale of $0.6''$ (half the size of the detector pixels) using an 
interlacing algorithm. In addition to regular sky maps, we also create
$A-B$ sky maps where all the odd numbered frames are multiplied by $-1$ before
mosaicking the image. This has the effect of removing the contributions 
of fixed celestial sources and leaving only instrumental noise 
(and photon shot noise, see Section \ref{sec:white}).

\begin{figure*}[t] 
   \centering
   \includegraphics[width=3.5in]{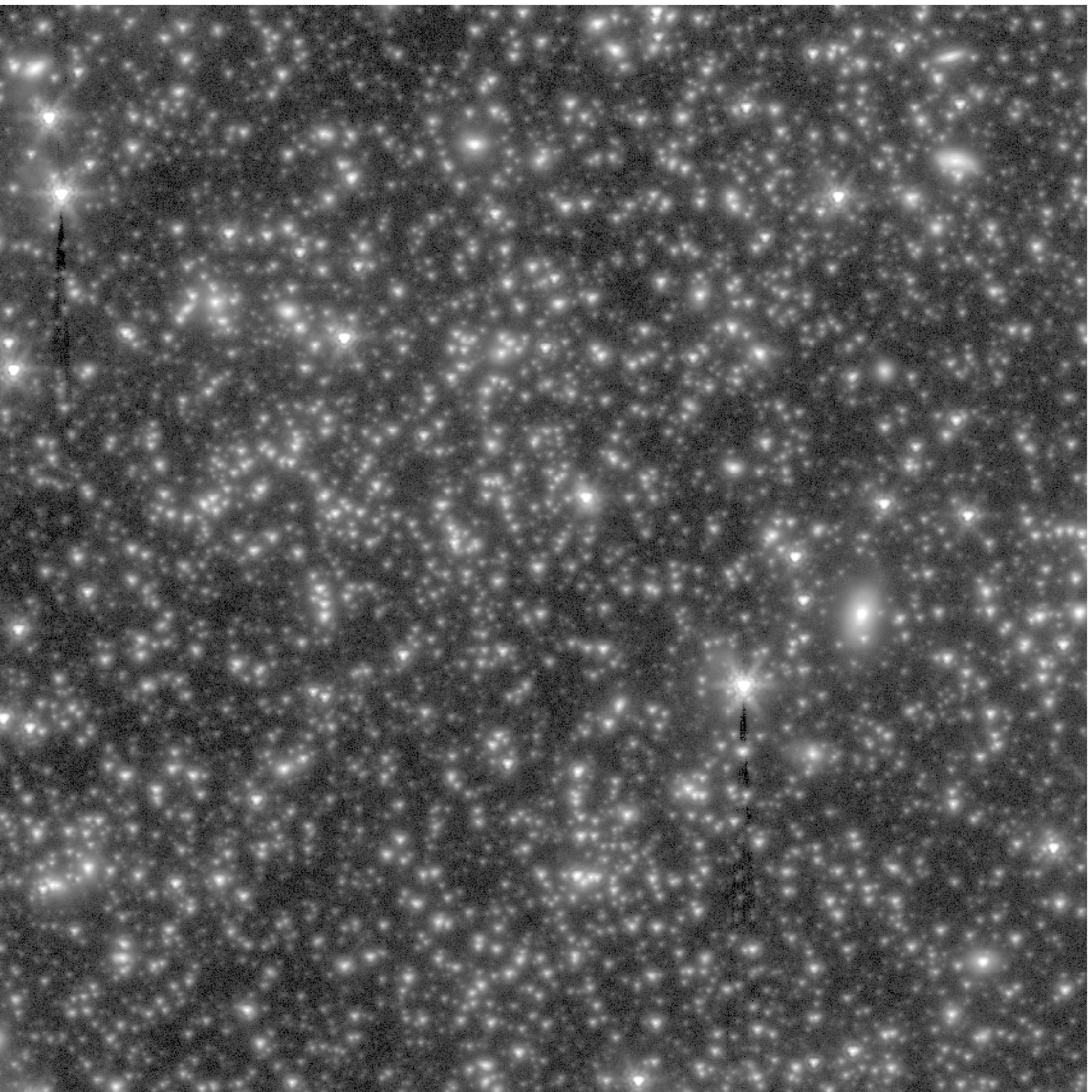}
   \includegraphics[width=3.5in]{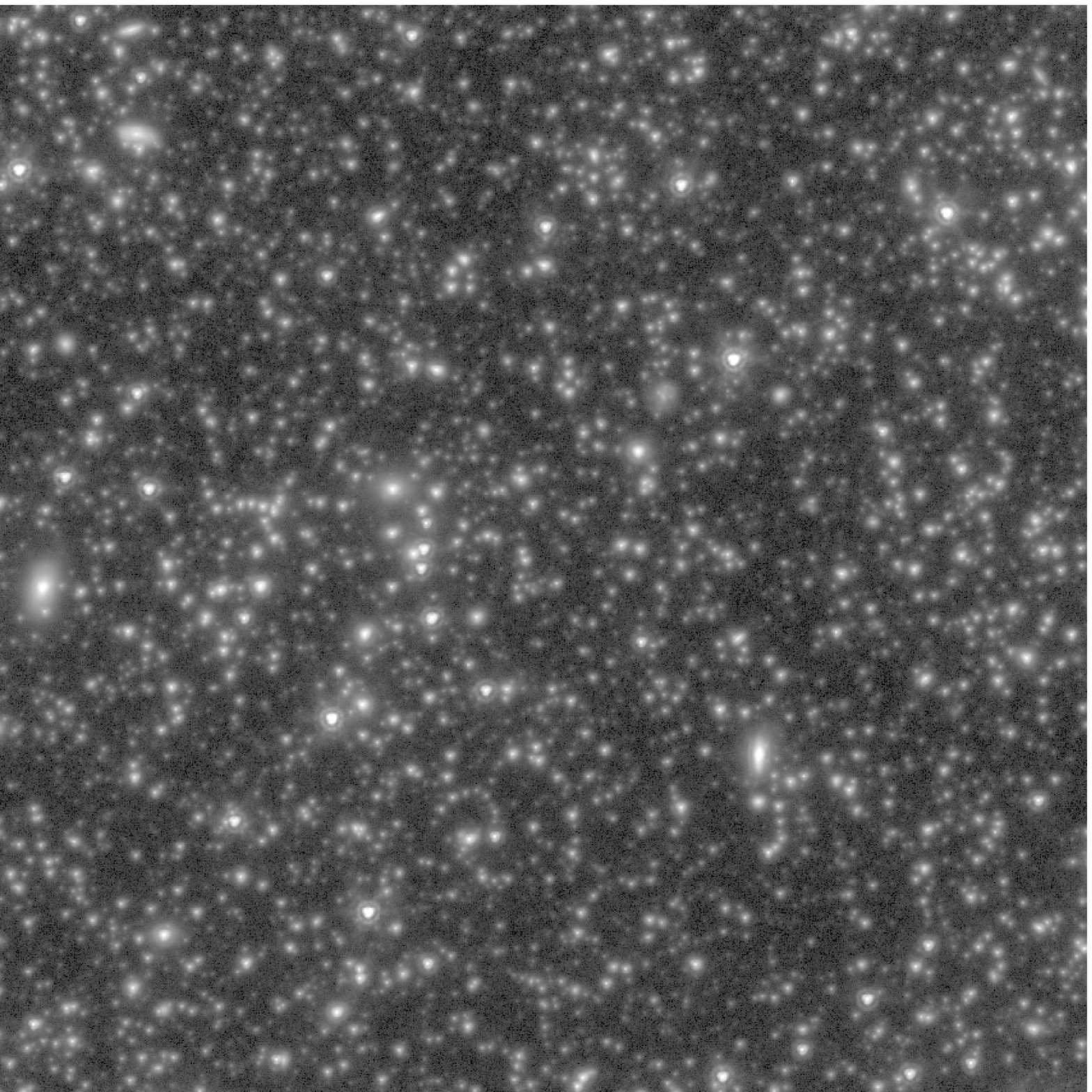} 
   \caption{\edit1{Combined five-epoch 3.6 $\mu$m (left) and 4.5 $\mu$m (right) 
   mosaics of the study fields, illustrating the source density and the 
   absence of bright stars and large galaxies.} Only the two brightest stars 
   in the 3.6 $\mu$m image show a small residual of incompletely corrected 
   ``column pull-down'' artifacts (dark vertical stripes). The images are logarithmically
   scaled on the range [$2.5\times10^{-4}$, 1.0] MJy sr$^{-1}$ after addition of
   an offset of 0.004 MJy sr$^{-1}$ to avoid logarithmic scaling of negative
   data values.
   The images are 900$\times$900 pixels or $9'\times9'$ in size (1 pixel = $0.6''$).
   Celestial north is at a position angle of $66\fdg4$ (counterclockwise 
   from vertical). There is only $\sim25\%$ overlap between the fields.
   \label{fig:image}}
\end{figure*}

Figure \ref{fig:image} shows the 3.6 and 4.5 $\micron$ images ($S^\alpha$ 
determined by self-calibration) of the 
combined 5 epochs of observations. The images are cropped to show only 
the roughly uniformly covered region that was used for power spectrum analysis.
Linear background gradients have been fitted and subtracted.

\begin{figure}[t] 
   \centering
   \includegraphics[width=3.5in]{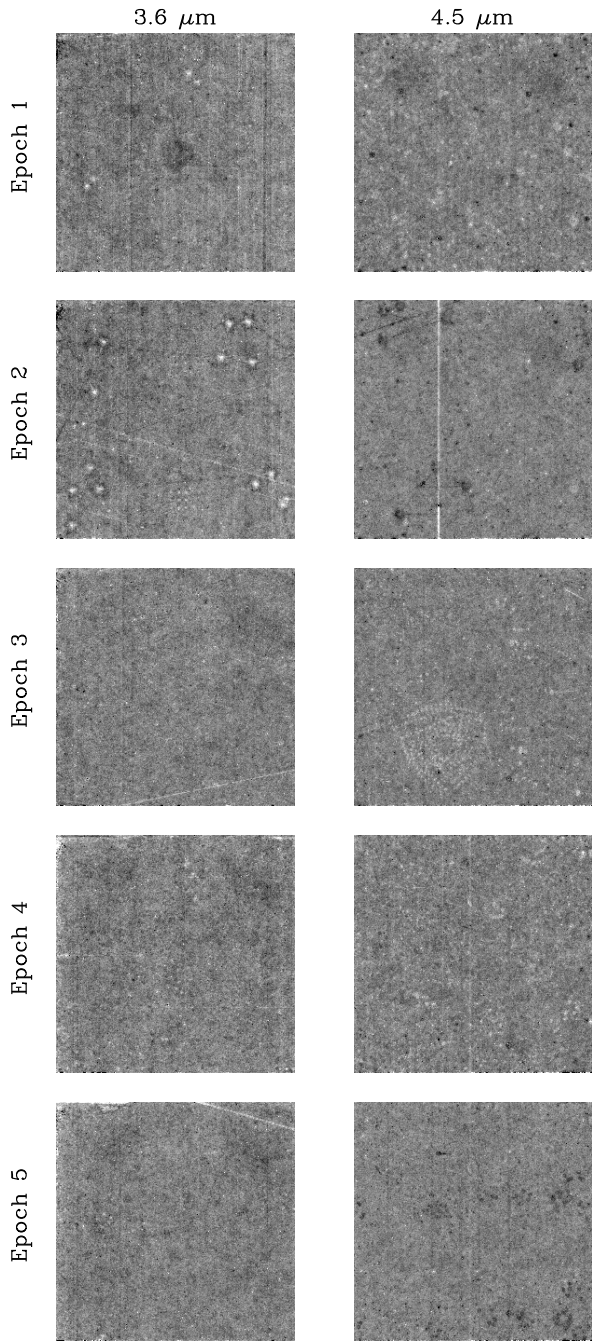} 
   \caption{Maps of the self-calibration $F^p$ offsets derived for each epoch.
   Different long-term latent images are present at each epoch. The effects of 
   residual straylight in the cBCD images are evident as diffuse dark patches 
   in the upper left and right corners of the later epochs at 3.6 $\micron$ 
   and the early epochs at 4.5 $\micron$. All images are on a linear stretch of 
   [$-0.015$, $+0.015$] MJy sr$^{-1}$.
   \label{fig:Fp}}
\end{figure}

Figure \ref{fig:Fp} displays the derived values of the detector offsets, 
$F^p$, for  3.6 and 4.5 $\mu$m at all 5 epochs. 
At all epochs there are different patterns of light and dark latent images,
which are very slowly decaying imprints of very bright stars as they were 
dithered across the detector in previous observations. In several cases
the tracks of bright stars that slewed across the detector between 
pointings can also be seen. 
Residual stray light in the cBCD frames is also revealed and removed by 
the self-calibration. 
Diffuse patches of stray light are created by
the zodiacal light (and the sum of all other backgrounds) 
in the upper left and upper right 
corners of both the 3.6 and 4.5 $\mu$m detectors. The BCD pipeline uses 
estimates of the expected brightness of the zodiacal light to model and 
remove this stray light component. These self-calibration results show that 
at 3.6 $\mu$m that process works well initially, when the zodiacal light is 
bright, but has an increasing error at later epochs as the elongation increases
and the zodiacal light becomes fainter. The opposite occurs at 4.5 $\mu$m, where
the standard correction works best at the later high-elongation epochs and less
well at the early epochs when the zodiacal light is brighter.
All these features visible in the $F^p$ maps are systematic effects that
are removed by the self-calibration of the data.

\begin{figure}[t] 
   \centering
   \includegraphics[width=3.5in]{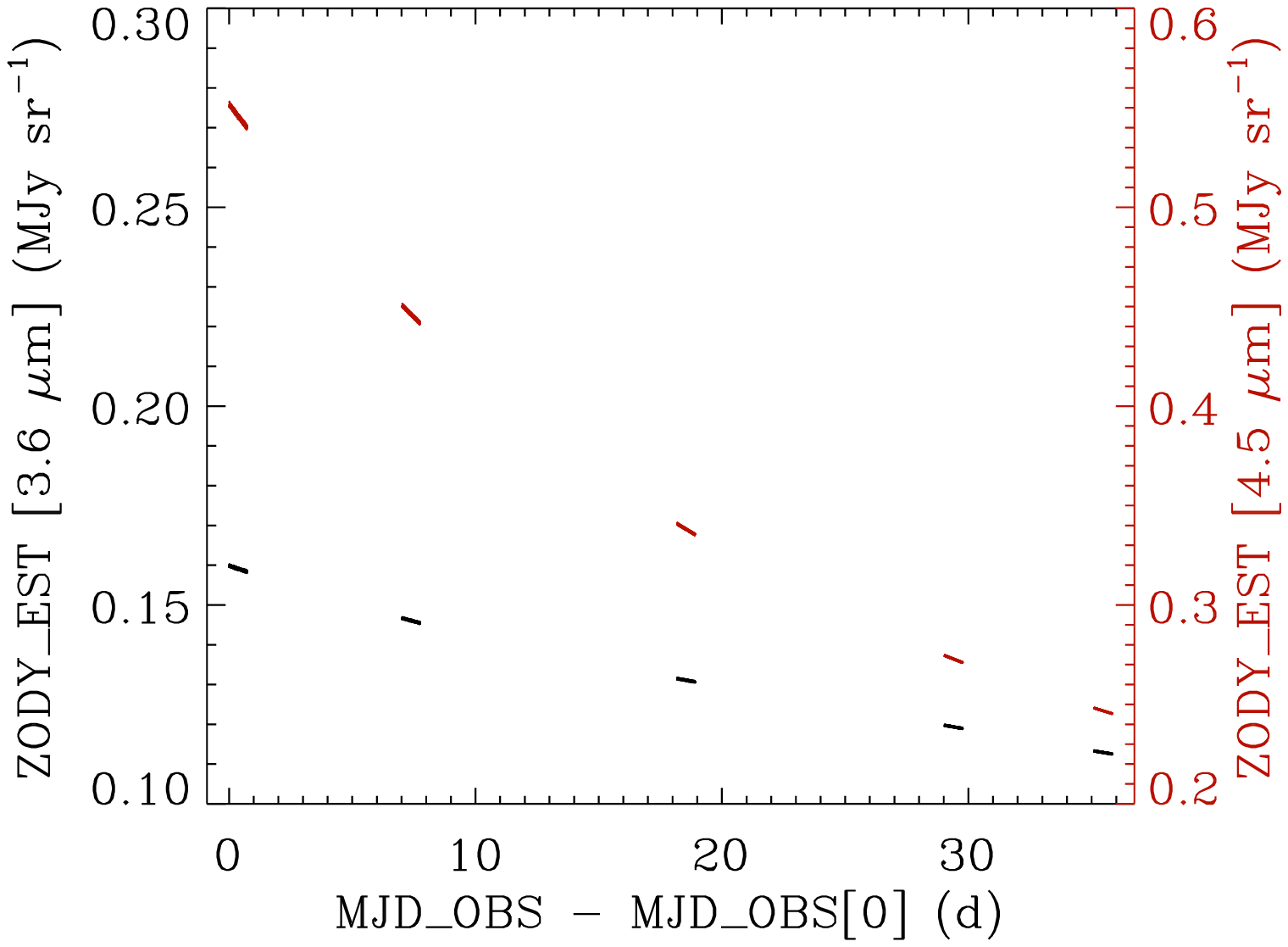}\\
   \vspace{0.1in}
   \includegraphics[width=3.5in]{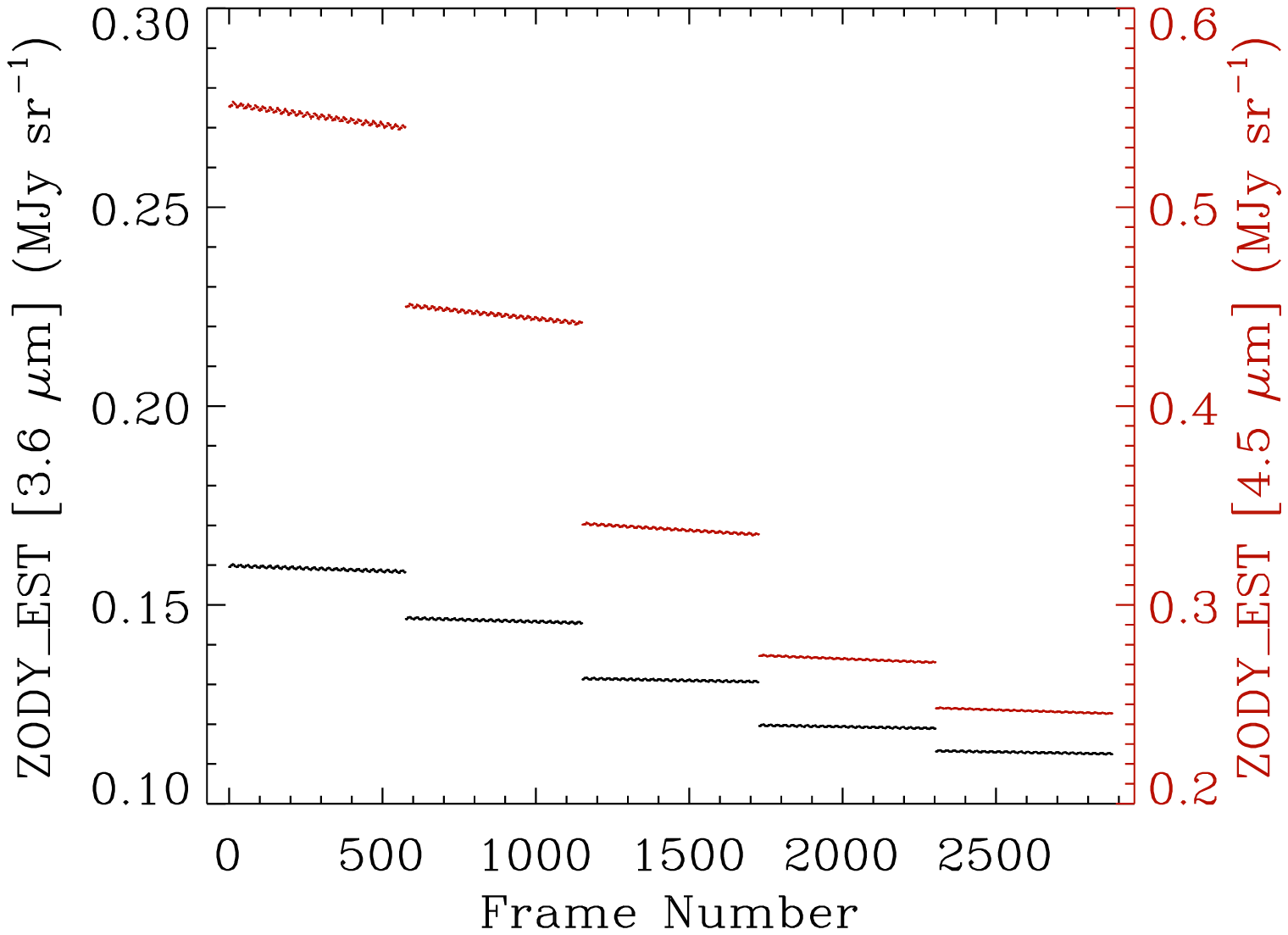} 
   \caption{The predicted zodiacal light intensity as given by the 
   ZODY\_EST keyword in the cBCD frames. The top plot shows the steady decline
   in intensity as a function of time. The fractional decrease at 4.5 $\mu$m
   is larger than that at 3.6 $\mu$m.
   Plotting simply as a function of 
   frame number (bottom) provides a slightly clearer look at the very small  
   oscillations in intensity that are caused by moving up and down the zodiacal
   light gradient at different pointings.
   \label{fig:zodi-est}}
\end{figure}

\begin{figure}[t] 
   \centering
   \includegraphics[width=3.5in]{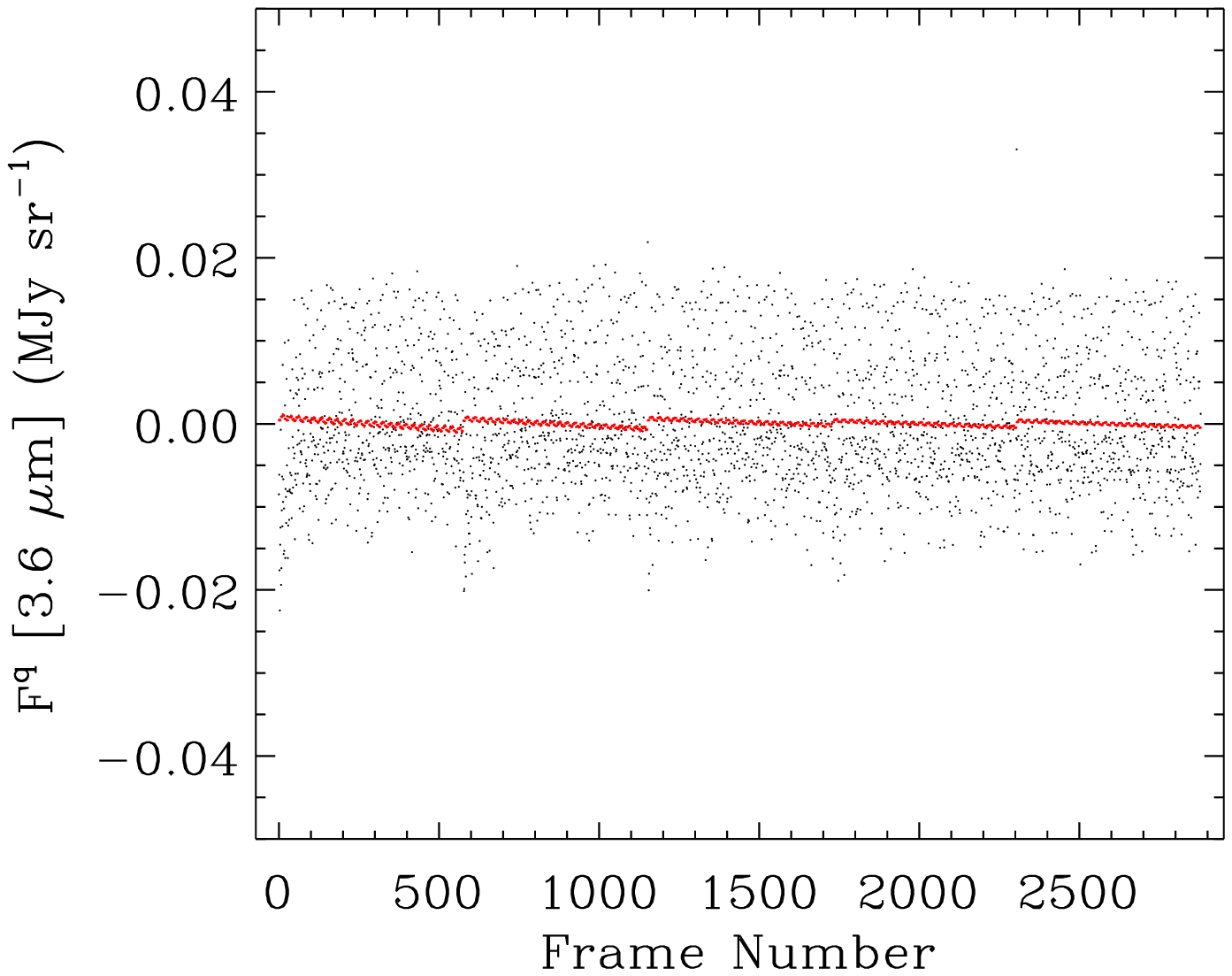}\\
   \vspace{0.1in}
   \includegraphics[width=3.5in]{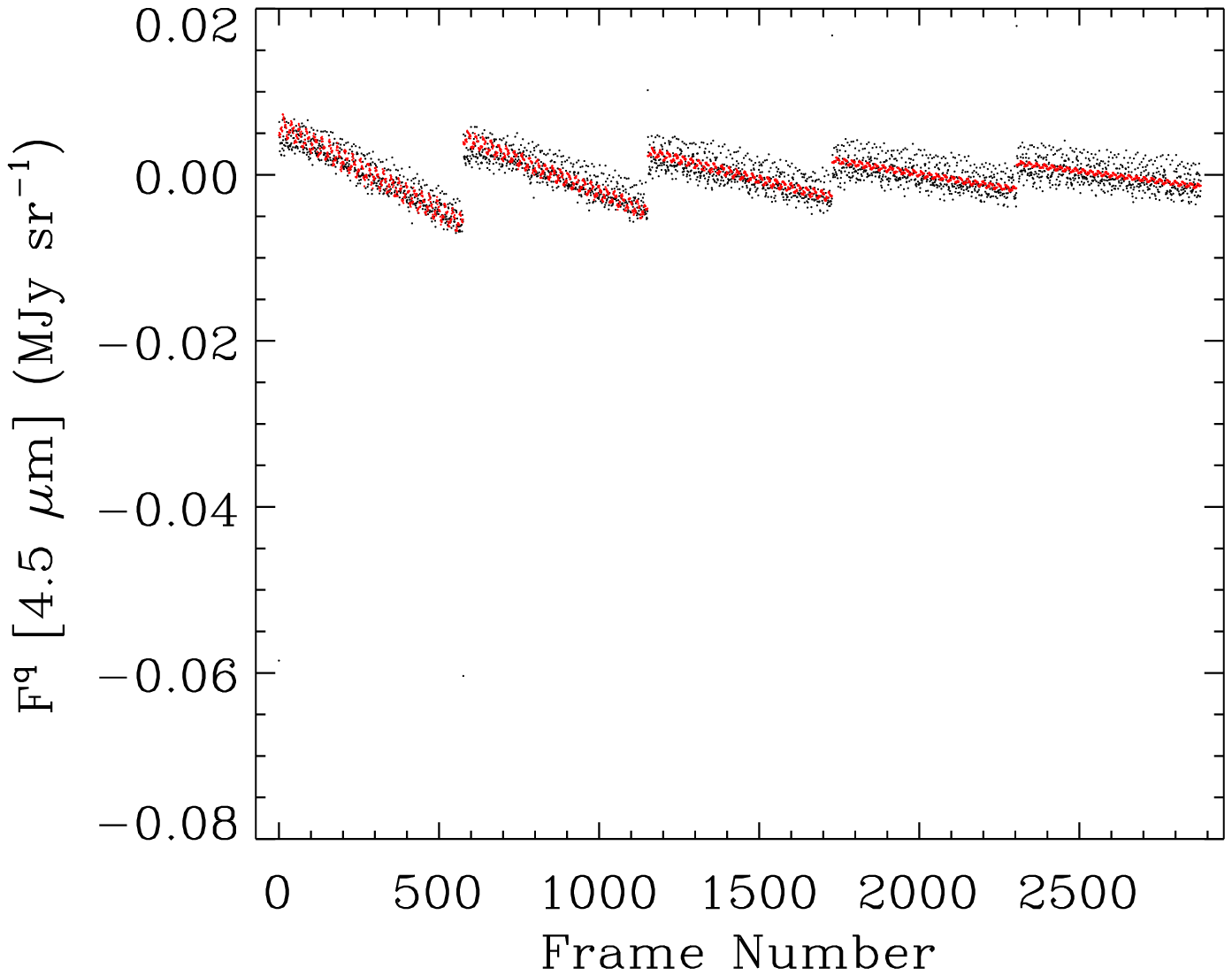} 
   \caption{Comparison between the self-calibration $F^q$ offsets (black dots)
   and the ZODY\_EST keyword values (red dots) after subtraction of the mean 
   values for each epoch. The lack of perfect correlation between these 
   reflects errors in the zodiacal light model and the presence of additional
   temporally variable effects in the data
   \edit1{(especially the first-frame effect at 3.6 $\mu$m, see the Appendix)},
   which are removed by $F^q$.
   \label{fig:Fq}}
\end{figure}

Figure \ref{fig:zodi-est} shows expected general trend of the zodiacal light intensity as a 
function of time, as estimated by the {\it Spitzer} foreground 
model\footnote{\burl{http://ssc.spitzer.caltech.edu/warmmission/propkit/som/bg/background.pdf}}
and given
as the ZODY\_EST keyword in the header of each frame. Note the 4.5 $\micron$
intensity is a stronger function of time or elongation than 
the 3.6 $\micron$ intensity, indicating that the color of the zodiacal light
is not expected to be constant.
The self-calibration model applied in Equation (\ref{eq:selfcal}) assumes a 
fixed sky intensity. Thus any variations in brightness due to changing 
zodiacal light intensity across the span of the data 
being self-calibrated, or a single epoch,
are absorbed by the variable offset term, $F^q$.
Figure \ref{fig:Fq} shows that $F^q$ varies by a much larger amount than the 
expected zodiacal light trends at 3.6~$\mu$m, but is similar to the 
expected zodiacal light trends at 4.5~$\mu$m. In the Appendix we show 
that the differences are real and represent corrections for transient 
instrumental effects (not fully corrected in the BCD pipeline), and 
residual linear gradients across the field.

\subsection{Source Subtraction and Masking}
For measurement of the power spectra of the background, resolved sources
need to be masked or modeled and subtracted from the images. As in 
prior studies, we subtract the flux from sources above the noise level 
using an iterative algorithm described by \cite{arendt:2010a}. 
The iterations are halted when the skewness of the intensity 
distribution of the remaining pixels is zero 
(independently for each epoch).
Because the removal is imperfect, the image is also masked using a mask
derived from a surface brightness threshold in the original images 
for all epochs combined.
The masking threshold can equivalently be expressed as a specified
surface brightness, a specified maximum outlier in the distribution
of surface brightness of unmasked pixels, or a specified fraction
of area masked. In this case we chose the last constraint, limiting
the masked out regions at both wavelengths 
to 25\% of the image, leaving 75\% of the image remaining.
At 3.6 and 4.5~$\mu$m, this limit corresponds surface brightness 
thresholds of ${\rm max}(I_\nu)-{\rm mean}(I_\nu) = 0.0057$ and 0.0044 MJy~sr$^{-1}$,
and to maximum outliers of 
${\rm max}(I_\nu)/\sigma_{I_\nu} = 3.2$ and 2.6 respectively.

\section{Power Spectra} \label{sec:power}

\begin{figure*}[p] 
   \begin{center}
   \includegraphics[height=8.5in]{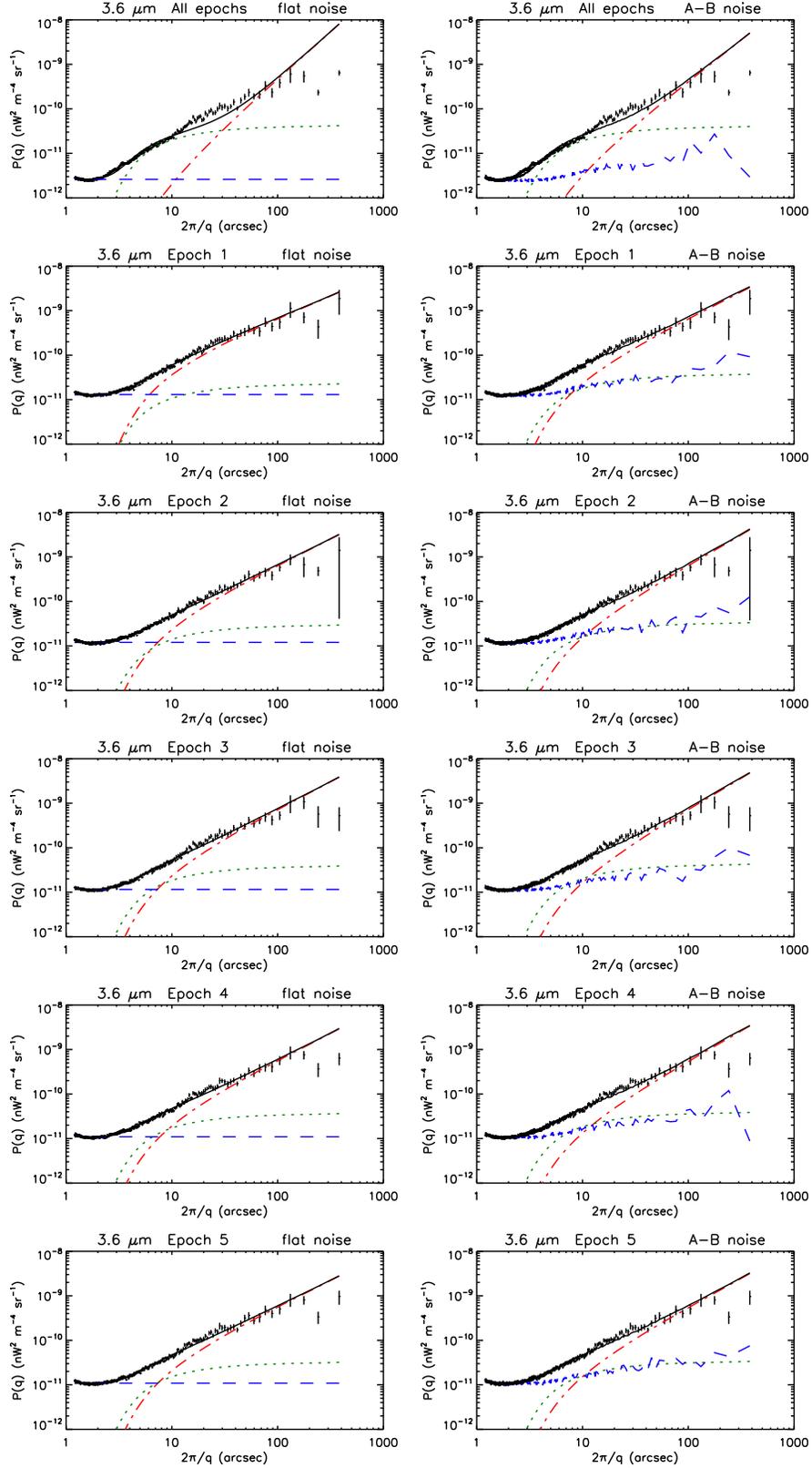}
   \end{center}
   \caption{Fits to the 3.6 $\micron$ power spectra for all epochs combined, and for each of 
   the 5 separate epochs.
   The left column shows fits as characterized by Equation (\ref{eq:model1}): \edit1{(a) flat white noise
   components (blue dashed line), (b) a flat shot noise component convolved with the PRF (green dotted line), and (c) a power
   law component, also convolved with the PRF (red dot-dashed line). The black solid line indicates the sum of
   these three components.} The right column shows the fits as characterized
   by Equation (\ref{eq:model2}), where the measured (A-B)/2 noise takes the place of the flat instrument
   noise component. 
   \label{fig:fit_plots1}}
\end{figure*}

\begin{figure*}[p] 
   \begin{center}
   \includegraphics[height=8.5in]{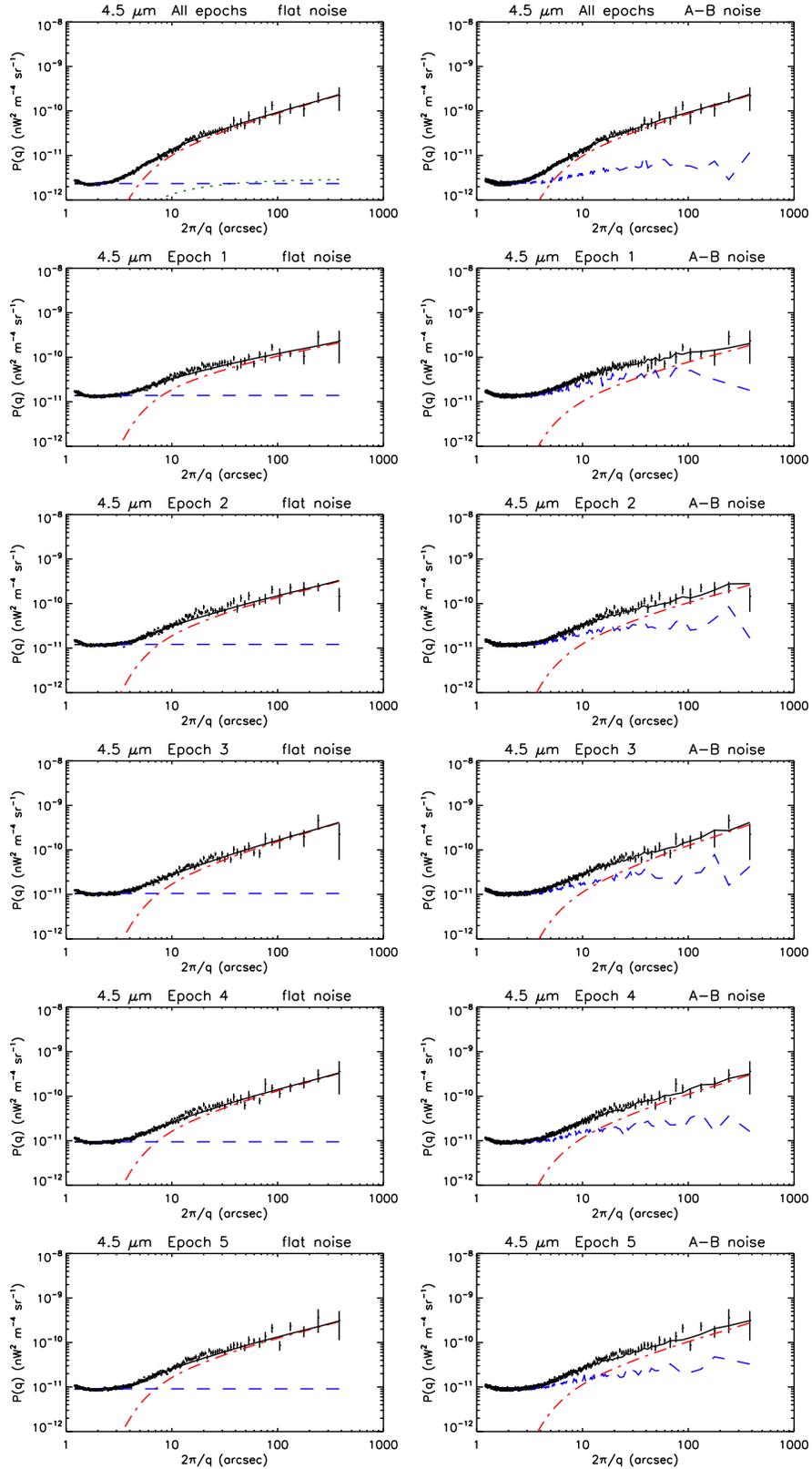}
   \caption{Same as Fig. \ref{fig:fit_plots1}, except for 4.5 $\micron$.
   \edit1{Note that a sky shot noise component is included in all of these 
   fits, but in most cases its preferred amplitude is 0.} 
   \label{fig:fit_plots2}}
   \end{center}
\end{figure*}

The power spectra of the source-subtracted images are calculated as 
described by \cite{arendt:2010a}. The power 
on the horizontal ($u=0$) and vertical ($v=0$) 
axes in the Fourier domain is omitted when averaging the power in bins at 
different angular scales, $2\pi/q$. This makes the results less 
susceptible to certain systematic errors (such as the residual
column pull-down seen at 3.6 $\mu$m in Figure \ref{fig:image}), but limits the maximum 
angular scale to $382'' = 540''/\sqrt2$ instead of the full $540''$
size of the field. 
\edit1{The uncertainties assigned to each binned point in the power 
spectra are the standard deviations of all measurements contributing
to each bin.}
The resulting power spectra, for the five epochs combined,
and for each epoch individually are plotted in 
Figures \ref{fig:fit_plots1} and \ref{fig:fit_plots2}.

The power spectra are characterized as in \cite{arendt:2010a} by fitting 
a combination of 3 simple components:
\begin{equation} \label{eq:model1}
P(q) = a_0(2\pi/q/100'')^{a_1}\ P_{\rm PRF}(q) + a_2\ P_{\rm PRF}(q) + a_3
\end{equation}
where the parameters $a_0$ and $a_1$ are the amplitude and index of a 
power-law component that is modulated by the instrument beam or point response
function, $P_{\rm PRF}(q)$, the parameter $a_2$ is the amplitude of the sky shot 
noise: a white shot-noise component that is also modulated by the beam (e.g. Poisson 
variation in the number of faint unresolvable sources in the beam at each location), 
and $a_3$ is
the amplitude of a white (shot noise) component that is not modulated by the beam.
As an alternate characterization, we also fit:
\begin{equation} \label{eq:model2}
P(q) = b_0(2\pi/q/100'')^{b_1}\ P_{\rm PRF}(q) + b_2\ P_{\rm PRF}(q) + P_{A-B}(q)
\end{equation}
where the white noise component is replaced by the measured $A-B$ power 
spectrum with no rescaling allowed.
These characterizations of the power spectra are overplotted in 
Figures \ref{fig:fit_plots1} and \ref{fig:fit_plots2} and are
tabulated in Table~\ref{tab:fit_params}.

\edit1{The values of reduced chi squared ($\chi^2_\nu$) in Table 2 are calculated for
each fit using the full data set, i.e. $\nu = 445$ or 446 degrees of freedom for 
Equations (2) or (3). However, for the 3.6 $\mu$m power spectra, we strongly 
deweight the two data points at $2\pi/q > 200''$ because the estimated uncertainties 
are clearly inconsistent with the behavior of these data points.
The absolute values of $\chi^2_\nu$ are poor, indicating that the uncertainties 
may be underestimated or these models do not accurately reflect the power spectra.
However, the overall aim of these fits is not to validate a particular physical 
model of the power spectra, but to reduce the power spectra to a small number of 
parameters that can be easily compared between power spectra (as we do below), 
and which provide a convenient approximation to the power spectra.}

We note that one could apply a more physical model for 
characterization of the large-scale component of the power spectra,
e.g. a $\Lambda$CDM template. However, a simple power law provides
a sufficient approximation for a wide possibility of origins, 
given the angular scales ($\lesssim400''$) and quality of the data analyzed here. 

\clearpage
\onecolumngrid
\begin{deluxetable}{llDDDDDDDDD}
\tabletypesize{\scriptsize}
\tablewidth{0pt}
\tablecaption{Power Spectrum Parameters\label{tab:fit_params}}
\tablehead{
\colhead{$\lambda$ ($\micron$)} &
\colhead{Epoch} &
\twocolhead{$10^{11}a_0$} &
\twocolhead{$a_1$} &
\twocolhead{$10^{11}a_2$} &
\twocolhead{$10^{11}a_3$} &
\twocolhead{$\chi^2_{\nu}$} &
\twocolhead{$10^{11}b_0$} &
\twocolhead{$b_1$} &
\twocolhead{$10^{11}b_2$} &
\twocolhead{$\chi^2_{\nu}$}
}
\decimals
\startdata
3.6 & All epochs & 50.24$\pm$ 2.82 &  2.07$\pm$ 0.06 &  4.19$\pm$ 0.04 &  0.26$\pm$ 0.001 & 4.47 & 43.76$\pm$ 2.90 &  1.82$\pm$ 0.06 &  4.01$\pm$ 0.05 & 5.92 \\
3.6 & Epoch 1    & 70.58$\pm$ 4.23 &  0.97$\pm$ 0.05 &  2.26$\pm$ 0.39 &  1.30$\pm$ 0.003 & 2.73 & 70.28$\pm$ 4.65 &  1.16$\pm$ 0.05 &  3.73$\pm$ 0.28 & 2.48 \\
3.6 & Epoch 2    & 68.39$\pm$ 4.21 &  1.15$\pm$ 0.05 &  2.95$\pm$ 0.25 &  1.20$\pm$ 0.002 & 2.99 & 68.90$\pm$ 4.60 &  1.33$\pm$ 0.06 &  3.33$\pm$ 0.20 & 2.76 \\
3.6 & Epoch 3    & 77.19$\pm$ 4.73 &  1.20$\pm$ 0.05 &  3.89$\pm$ 0.24 &  1.16$\pm$ 0.002 & 2.44 & 77.67$\pm$ 5.09 &  1.35$\pm$ 0.05 &  4.26$\pm$ 0.20 & 2.90 \\
3.6 & Epoch 4    & 60.29$\pm$ 3.60 &  1.18$\pm$ 0.05 &  3.63$\pm$ 0.22 &  1.09$\pm$ 0.002 & 2.67 & 58.09$\pm$ 3.75 &  1.32$\pm$ 0.06 &  3.86$\pm$ 0.18 & 2.52 \\
3.6 & Epoch 5    & 60.30$\pm$ 3.65 &  1.14$\pm$ 0.05 &  3.20$\pm$ 0.23 &  1.09$\pm$ 0.002 & 2.54 & 58.36$\pm$ 3.82 &  1.26$\pm$ 0.05 &  3.36$\pm$ 0.20 & 3.38 \\
\hline
4.5 & All epochs &  9.61$\pm$ 0.19 &  0.65$\pm$ 0.02 &  0.29$\pm$ 0.10 &  0.23$\pm$ 0.001 & 3.55 &  9.44$\pm$ 0.20 &  0.65$\pm$ 0.02 &  0.06$\pm$ 0.11 & 7.25 \\
4.5 & Epoch 1    & 11.59$\pm$ 0.63 &  0.46$\pm$ 0.04 &  0.00$\pm$ 0.57 &  1.39$\pm$ 0.003 & 5.21 &  8.50$\pm$ 0.52 &  0.59$\pm$ 0.06 &  0.00$\pm$ 0.40 & 4.56 \\
4.5 & Epoch 2    & 14.94$\pm$ 0.51 &  0.56$\pm$ 0.03 &  0.00$\pm$ 0.34 &  1.21$\pm$ 0.003 & 5.15 & 11.38$\pm$ 0.53 &  0.63$\pm$ 0.04 &  0.00$\pm$ 0.31 & 4.51 \\
4.5 & Epoch 3    & 16.76$\pm$ 0.40 &  0.66$\pm$ 0.03 &  0.00$\pm$ 0.25 &  1.05$\pm$ 0.002 & 4.27 & 13.62$\pm$ 0.41 &  0.75$\pm$ 0.04 &  0.00$\pm$ 0.22 & 4.24 \\
4.5 & Epoch 4    & 14.37$\pm$ 0.36 &  0.61$\pm$ 0.03 &  0.00$\pm$ 0.28 &  0.95$\pm$ 0.002 & 4.11 & 12.11$\pm$ 0.37 &  0.68$\pm$ 0.04 &  0.00$\pm$ 0.25 & 3.57 \\
4.5 & Epoch 5    & 13.52$\pm$ 0.38 &  0.58$\pm$ 0.03 &  0.00$\pm$ 0.30 &  0.91$\pm$ 0.002 & 3.81 & 11.36$\pm$ 0.38 &  0.66$\pm$ 0.04 &  0.00$\pm$ 0.26 & 4.29 \\
\hline
4.5 & All epochs &  8.34$\pm$ 0.17 & 1.0 &  1.35$\pm$ 0.02 &  0.23$\pm$ 0.000 & 3.93 &  8.15$\pm$ 0.17 & 1.0 &  1.09$\pm$ 0.02 & 7.91 \\
4.5 & Epoch 1    &  7.46$\pm$ 0.19 & 1.0 &  2.88$\pm$ 0.06 &  1.38$\pm$ 0.003 & 6.03 &  6.46$\pm$ 0.21 & 1.0 &  1.29$\pm$ 0.07 & 4.81 \\
4.5 & Epoch 2    &  9.22$\pm$ 0.22 & 1.0 &  2.57$\pm$ 0.06 &  1.20$\pm$ 0.002 & 6.30 &  7.76$\pm$ 0.24 & 1.0 &  1.44$\pm$ 0.07 & 5.00 \\
4.5 & Epoch 3    & 14.32$\pm$ 0.31 & 1.0 &  1.81$\pm$ 0.05 &  1.04$\pm$ 0.002 & 4.92 & 12.24$\pm$ 0.32 & 1.0 &  0.87$\pm$ 0.06 & 4.47 \\
4.5 & Epoch 4    & 12.02$\pm$ 0.24 & 1.0 &  1.92$\pm$ 0.05 &  0.94$\pm$ 0.002 & 4.82 & 10.51$\pm$ 0.25 & 1.0 &  1.18$\pm$ 0.05 & 3.91 \\
4.5 & Epoch 5    & 11.09$\pm$ 0.26 & 1.0 &  1.99$\pm$ 0.05 &  0.90$\pm$ 0.002 & 4.62 &  9.84$\pm$ 0.26 & 1.0 &  1.17$\pm$ 0.05 & 4.75 \\
\enddata
\tablecomments{Units for $a_0$, $a_2$, $a_3$, $b_0$, and $b_2$ are nW$^2$ m$^{-4}$ sr$^{-1}$.}
\end{deluxetable}
\twocolumngrid

\subsection{White Noise Component} \label{sec:white}
The white noise component of the power spectrum, $a_3$ in Equation (\ref{eq:model1}),
includes instrumental noise, but it also includes the photon shot 
noise from celestial sources. In particular, the zodiacal light is 
the dominant brightness component. Because 
it is an approximately uniform source 
of emission, the power spectrum of its photon shot noise is not modulated
by the beam. As a noise term it also does not cancel out in the 
construction of the $A-B$ difference images, and therefore the
power spectra of those images also include the photon shot noise of the 
zodiacal \edit1{light.} 

\edit1{The} slight rise in the white noise component 
at the smallest angular scales \edit1{is} an artifact of mapping \edit1{the} data, sampled on $1\farcs2$ detector pixels,
onto a parallel sky map with $0\farcs6$ pixels. \edit1{Given our interlacing 
mapping algorithm, a slight mismatch in the mean level of any single frame
will insert power into the map at the Nyquist frequency of the $0\farcs6$ 
pixels of the sky map, i.e. at $1\farcs2$. The multiplication of the image
by a mask then convolves this power in the Fourier domain, spreading it 
to larger angular scales. Thus the shape of the turn up at the 
smallest spatial scales is related to the masking, and the amplitude is 
related to the size of the frame-to-frame background errors. Mapping on 
a grid that is not so well aligned with detector orientation can mitigate
the effect somewhat.}
For example, if the sky map is 
generated on grid that is rotated by $45\arcdeg$ to the detector orientation, 
then the white noise component does appear flat. \edit1{Because this rise cannot be 
well fit with flat white noise, more than half the 
contribution to $\chi^2_\nu$ comes from $2\pi/q < 2''$.}

\begin{figure}[t] 
\centering
\includegraphics[height=2.5in]{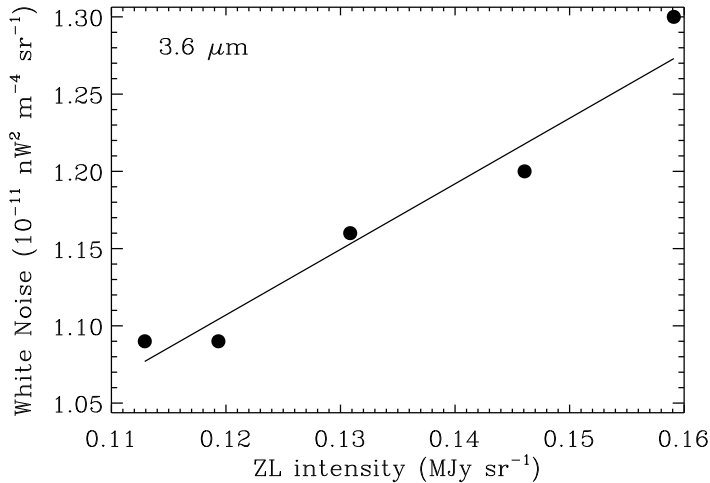}\\ 
\vspace{0.1in}
\includegraphics[height=2.5in]{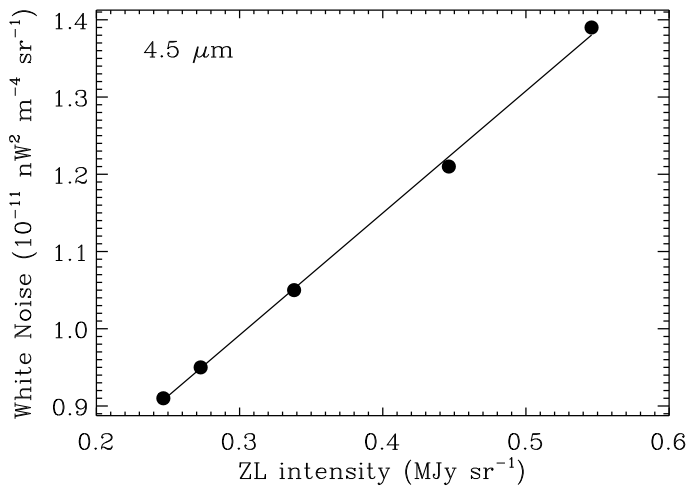} 
\caption{Correlation between the white noise level ($a_3$) and the 
zodiacal light intensity across the 5 epochs shows that roughly half of the 
white noise is correlated with the zodiacal light. The formal uncertainties in 
the white noise values are smaller than the plotted symbols (see Table 
\ref{tab:fit_params}) \edit1{The Kendall's $\tau$ rank correlations 
(bounded on the interval [-1,1]) for these quantities at 3.6 and 
4.5 $\mu$m are 0.95 and 1.0. The probabilities of finding these 
values of $|\tau|$ (or larger) given uncorrelated data 
are 0.02 and 0.01.}
\label{fig:whitenoise}}
\end{figure}

Figure \ref{fig:whitenoise} shows the very strong correlation 
between the zodiacal light intensity and the measured level of the white 
noise power spectrum. Extrapolation to zero intensity of the zodiacal light
indicates that white noise power 
levels in the absence of zodiacal light would be
$\edit1{5.7}\times10^{-12}$ and $\edit1{5.2}\times10^{-12}$ nW$^2$ m$^{-4}$ sr$^{-1}$ at
3.6 and 4.5 $\mu$m respectively.
\edit1{Comparison to the measured white noise power ($a_3$ in 
Table \ref{tab:fit_params}) indicates that photon shot noise of the zodiacal 
light contributes $\sim40-60\%$ of the amplitude of this component, 
depending on the epoch of the observations.}

\subsection{Sky Shot Noise and Power Law \edit1{Components}}

The sky shot noise component, characterized by $a_2$ or $b_2$, is not an essential 
component for fitting the observed power spectra for the five individual epochs.
These observations are so shallow that the flat white noise component 
can dominate to sufficiently large scales where the power law component takes over.
At 4.5 $\mu$m, the best fits are obtained without a sky shot noise 
component, though this requires shallower power law indices, 
$a_1$ and $b_1$, than previously found \citep{arendt:2010a}. 
This is likely caused by an inability to distinguish separate sky shot 
noise and power law components in these shallow data. 
Constraining the power law index to be $a_1 = b_1 = 1.0$ 
does result in a weak sky shot noise component that 
constitutes $\sim15\%$ of the power at $100''$, but 
the constraint produces a poorer fit at scales of $\sim20''$. 
Results of this constrained fit 
are listed in the last section of Table~\ref{tab:fit_params}.
At 3.6 $\mu$m there is no similar motivation for a constrained fit,
as the best fits find non-zero sky shot noise components ($a_2$ 
and $b_2$) and power law slopes ($a_1$ and $b_1$) that are similar
to those previously derived \citep{arendt:2010a}.
\edit1{Conversely, omitting the sky shot noise component
($a_2$ and $b_2$) at 3.6~$\mu$m causes the power law component
to flatten, to better fit the spectrum at $2\pi/q \lesssim 10''$.
For the all-epochs-combined power spectrum, the power law 
flattens completely, ($a_1 \sim 0$ and $b_1 \sim 0$), to
become the sky shot noise component, leaving the rising power
at $2\pi/q \gtrsim 10''$ poorly fit.} 

\begin{figure}[t] 
\centering
\includegraphics[height=2.5in]{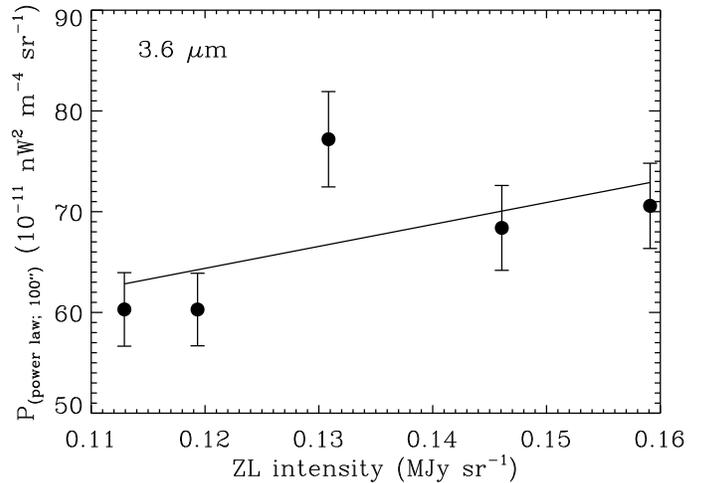}\\ 
\vspace{0.1in}
\includegraphics[height=2.5in]{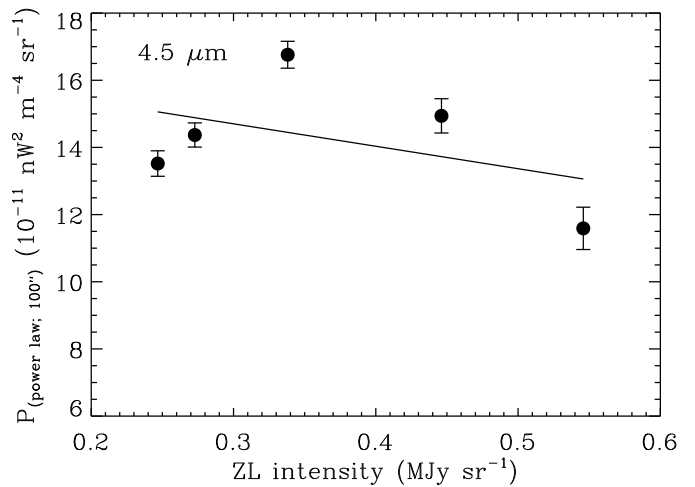} 
\caption{There is no significant correlation
between the amplitude of the power law component 
($a_0$) and the zodiacal light intensity across the 5 epochs 
at either 3.6 $\mu$m (top) or 4.5 $\mu$m (bottom).
However, the results shown here are sensitive to the choice
of model depth (See Appendix).
\edit1{The Kendall's $\tau$ rank correlations 
(bounded on the interval [-1,1]) for these quantities at 3.6 and 
4.5 $\mu$m are 0.4 and 0.0. The probabilities of finding these 
values of $|\tau|$ (or larger) given uncorrelated data 
are 0.33 and 1.0.}
\label{fig:powernoise}}
\end{figure}

The amplitudes, $a_0$ and $b_0$, of the power law fitting the large scale power 
at 4.5 $\mu$m are much lower than at 3.6 $\mu$m, unlike prior
results where powers were comparable \citep{arendt:2010a}. 
Figure \ref{fig:powernoise} shows \edit1{no} correlation between 
the amplitude of the power law component and the intensity 
of the zodiacal light.

\section{Discussion} \label{sec:discuss}

\begin{figure}[t] 
\centering
\includegraphics[height=2.25in]{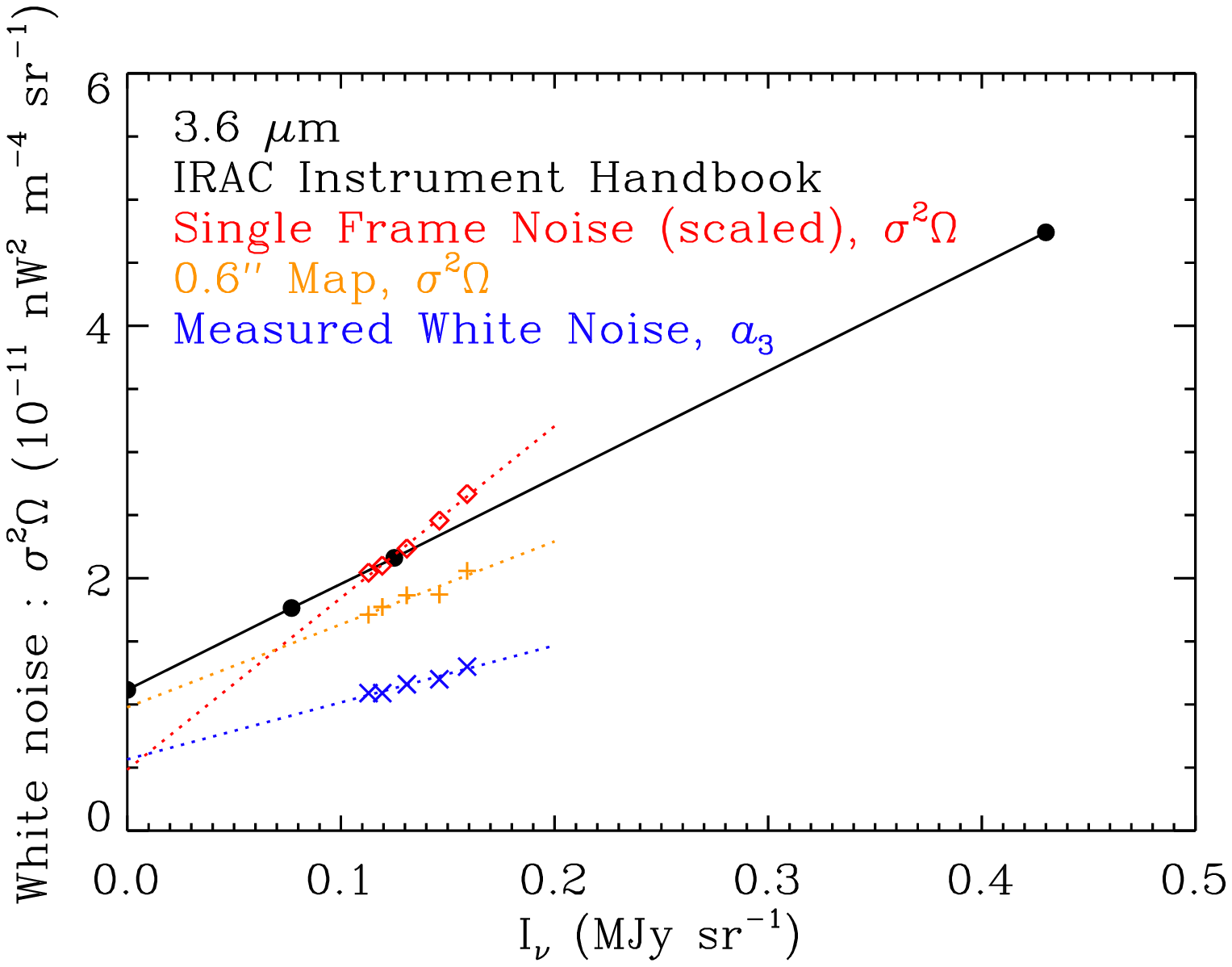}\\ 
\includegraphics[height=2.25in]{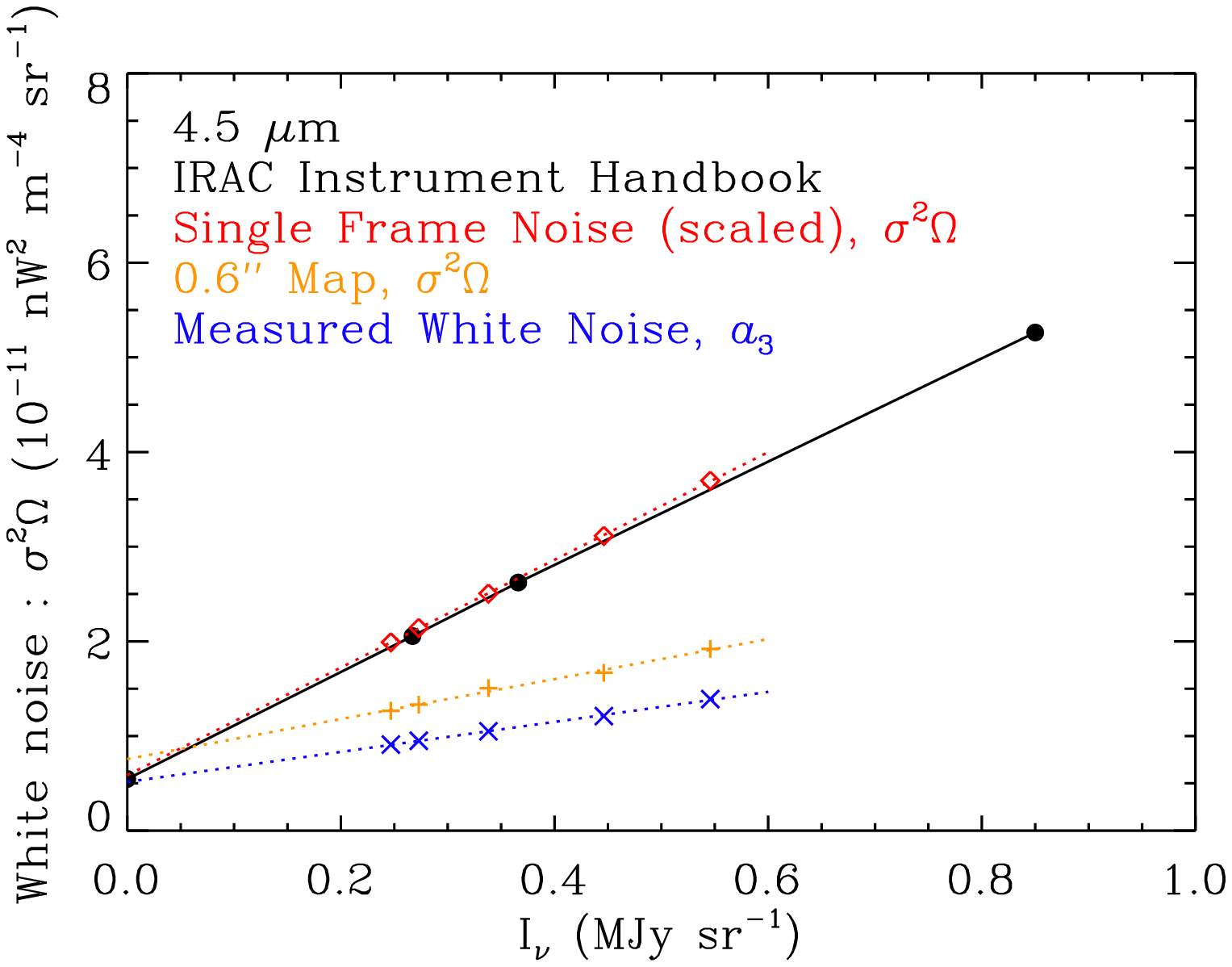} 
\caption{\edit1{Comparison of white noise predictions
and measurements as a function of background intensity. 
The expected white noise levels calculated
from the IRAC Instrument Handbook using ``High'', ``Medium'', 
and ``Low'' background levels, and with the background set to 0.0,
are shown by the black circles. The mean white noise levels derived from 
the measured standard deviation of individual frames at each of 
the 5 epochs are shown by red diamonds. The mean white noise levels
derived from the measured standard deviation of the 
self-calibrated sky maps are shown by orange $+$ symbols.
The measured white noise levels obtained by fits to the power spectra 
are indicated by the blue $\times$ symbols. See text for explanation 
of consistencies and discrepancies.
}
\label{fig:whitenoise2}}
\end{figure}

\edit1{The correlation between the zodiacal light intensity and the 
white noise in the power spectra (Figure \ref{fig:whitenoise}) indicates
both a constant component (i.e. the intercept as the zodiacal light intensity 
goes to 0) and a component that is proportional
to the zodiacal light intensity. 
Instrumental noise terms (e.g. read noise and dark current),
and the photon shot noise from the mean intensity 
of unresolved Galactic and extragalactic backgrounds 
would contribute to this constant term.
In Figure \ref{fig:whitenoise2} we present a comparison between 
the measured white noise power as a function of zodiacal 
light intensity, and the expected noise levels.}

\edit1{Using the equations and parameters presented in Section 2.5 of the 
IRAC Instrument 
Handbook\footnote{\burl{http://irsa.ipac.caltech.edu/data/SPITZER/docs/irac/iracinstrumenthandbook/}}, 
we can evaluate the expected 1-$\sigma$ noise level
for extended emission for 100s frame time exposures during the 
IRAC warm mission. These numbers are listed in Table \ref{tab:noise} (Note 1),
and, after conversion to white noise power, $P = \sigma^2\Omega_{\rm{pix}}$ 
(Table \ref{tab:noise}, Note 5), are shown in Figure 
\ref{fig:whitenoise2}.
For direct comparison to these expected noise estimates, 
we calculated the mean of the 
standard deviations of all (576) individual exposures at each epoch. 
A robust procedure was used to exclude the effect of stars and other
statistical outliers. These numbers (Table \ref{tab:noise}, note 2 and note 6)
are in generally good agreement 
with the expected uncertainties, with the exception that at 3.6 $\mu$m
the measured standard deviations show more variation with zodiacal 
light intensity than expected. This discrepancy may indicate that the 
zodiacal light model predicts the correct mean intensity of the 
zodiacal light, but underestimates the modulation of the intensity
as a function of time. This effect is very similar to that observed at
the north ecliptic pole (NEP) by \cite{krick:2012}.}

\edit1{The standard deviation in mosaicked sky maps should be reduced by 
a factor of $1/\sqrt{N}$, where $N$ is the number of frames contributing
at each pixel (Table \ref{tab:noise}, note 3). 
In this case there are 144 frames per pointing, but 
interlaced mapping on $0\farcs6$ pixels reduces the per pixel coverage
by a factor of 4, so $N = 36$. The actual measured standard deviations 
of the self-calibrated mosaics (Table \ref{tab:noise}, note 4) are $\sim10\%$
smaller at 3.6 $\mu$m and $\sim24\%$ smaller at 4.5 $\mu$m. Converted to a power
(Table \ref{tab:noise}, note 7), 
these values are plotted as the orange $+$ symbols in Figure \ref{fig:whitenoise2}.
They show a similar trend to the white noise levels derived from the power
spectra ($a_3$, blue $\times$ symbols), but are biased upwards because they
average the power at all angular scales, rather than just fitting the 
minimal white noise at the smallest angular scales.
The roughly constant size of the bias is a further indication that
the large scale structure is independent to the zodiacal light intensity.
}

\edit1{Ideally, the white noise level extrapolated to a 
zodiacal light intensity of 0, could be interpreted in terms of
instrumental noise plus the photon shot noise from galactic and 
extragalactic backgrounds. However, systematic errors in 
the zodiacal light model, as clearly evident at 3.6 $\mu$m, 
will directly affect the intercept. 
This is the likely reason for the mismatch between the extrapolated 
intercepts for the measured 3.6 $\mu$m power (either from the fitted power spectra 
or from single frames) and the expected noise power as calculated
from the IRAC Instrument Handbook. However,} 
the ultimate origin of the white noise component is not very important 
for current CIB studies, which normally subtract any ``instrumental''
or $A-B$ noise term that is constructed to be independent of
fixed sources on the sky. 

The large-scale component in the source-subtracted CIB power spectrum
is the term that is of greatest cosmological interest. Prior studies agree
that there is power here in excess of that expected from \edit1{the} 
faint unresolved galaxies extrapolated from known galaxy populations
\citep{kashlinsky:2005,sullivan:2007,helgason:2012,cooray:2012}.
The lack of significant correlation between the large-scale power and the 
zodiacal light intensity (Figure \ref{fig:powernoise}) suggests that 
the zodiacal light is not influencing the large scale power.
Additionally, we note that while the zodiacal light intensity is 
greater at 4.5 $\mu$m than at 3.6 $\mu$m, the data exhibit 
weaker large scale power at the 4.5 $\mu$m than at 3.6 $\mu$m.
This also indicates that the zodiacal light is not the 
main source of the large scale power.

\onecolumngrid
\edit1{
\begin{deluxetable}{llDDDDDcc}
\tabletypesize{\scriptsize}
\tablewidth{0pt}
\tablecaption{White Noise Estimates and Measurements\label{tab:noise}}
\tablehead{
\colhead{$\lambda$ ($\micron$)} &
\colhead{Quantity} &
\twocolhead{Epoch 1} &
\twocolhead{Epoch 2} &
\twocolhead{Epoch 3} &
\twocolhead{Epoch 4} &
\twocolhead{Epoch 5} &
\colhead{$I_{\nu}$(zodi) = 0} & 
\colhead{Note}
}
\decimals
\startdata
3.6 & ZODY\_EST (MJy sr$^{-1}$) &  0.159 &  0.146 &  0.131 &  0.119 &  0.113 & 0.0 & \\
3.6 & $\sigma_{\rm{frame}}$ (MJy sr$^{-1}$) & 0.0119 & 0.0116 & 0.0113 & 0.0110 & 0.0109 & \nodata & 1\\
3.6 & $\sigma_{\rm{frame}}$ (MJy sr$^{-1}$) & 0.0124 & 0.0119 & 0.0113 & 0.0110 & 0.0108 & \nodata & 2\\
3.6 & $\sigma$ (MJy sr$^{-1}$) & 0.00207 & 0.00198 & 0.00189 & 0.00183 & 0.00181 & \nodata & 3\\
3.6 & $\sigma$ (MJy sr$^{-1}$) & 0.00181 & 0.00173 & 0.00173 & 0.00168 & 0.00165 & \nodata & 4\\
3.6 & $\sigma^2\Omega$ (10$^{-11}$ nW$^2$ m$^{-4}$ sr$^{-1}$) &   2.45 &   2.34 &   2.21 &   2.11 &   2.06 & 1.12 & 5\\
3.6 & $\sigma^2\Omega$ (10$^{-11}$ nW$^2$ m$^{-4}$ sr$^{-1}$) &   2.67 &   2.46 &   2.24 &   2.10 &   2.05 & 0.48$\pm$0.09 & 6\\
3.6 & $\sigma^2\Omega$ (10$^{-11}$ nW$^2$ m$^{-4}$ sr$^{-1}$) &   2.06 &   1.87 &   1.87 &   1.77 &   1.71 & 0.98$\pm$0.17 & 7\\
3.6 & $a_3$ (10$^{-11}$ nW$^2$ m$^{-4}$ sr$^{-1}$) &   1.30 &   1.20 &   1.16 &   1.09 &   1.09 & 0.57$\pm$0.07 & 8\\
\hline
4.5 & ZODY\_EST (MJy sr$^{-1}$) &  0.546 &  0.446 &  0.338 &  0.273 &  0.247 & 0.0 & \\
4.5 & $\sigma_{\rm{frame}}$ (MJy sr$^{-1}$) & 0.0183 & 0.0169 & 0.0152 & 0.0140 & 0.0135 & \nodata & 1\\
4.5 & $\sigma_{\rm{frame}}$ (MJy sr$^{-1}$) & 0.0186 & 0.0171 & 0.0153 & 0.0141 & 0.0136 & \nodata & 2\\
4.5 & $\sigma$ (MJy sr$^{-1}$) & 0.00310 & 0.00284 & 0.00255 & 0.00236 & 0.00227 & \nodata & 3\\
4.5 & $\sigma$ (MJy sr$^{-1}$) & 0.00223 & 0.00208 & 0.00198 & 0.00186 & 0.00181 & \nodata & 4\\
4.5 & $\sigma^2\Omega$ (10$^{-11}$ nW$^2$ m$^{-4}$ sr$^{-1}$) &   3.60 &   3.06 &   2.46 &   2.09 &   1.94 & 0.55 & 5\\
4.5 & $\sigma^2\Omega$ (10$^{-11}$ nW$^2$ m$^{-4}$ sr$^{-1}$) &   3.70 &   3.11 &   2.51 &   2.14 &   1.99 & 0.59$\pm$0.01 & 6\\
4.5 & $\sigma^2\Omega$ (10$^{-11}$ nW$^2$ m$^{-4}$ sr$^{-1}$) &   1.92 &   1.67 &   1.51 &   1.33 &   1.27 & 0.76$\pm$0.04 & 7\\
4.5 & $a_3$ (10$^{-11}$ nW$^2$ m$^{-4}$ sr$^{-1}$) &   1.39 &   1.21 &   1.05 &   0.95 &   0.91 & 0.52$\pm$0.01 & 8\\
\enddata
\tablecomments{(1) Expected extended source surface brightness sensitivity for single 100s frames, 
as calculated from equations in Section 2.5 of the IRAC Instrument Handbook. 
(2) Measured average standard deviation (outliers excluded) of single frames. 
(3) Measured average standard deviation (outliers excluded) of single frames scaled by of $1/\sqrt{N}$, 
(4) Average standard deviation (outliers excluded) of self-calibrated sky map ($0\farcs6$ pixels).
(5) White noise power derived from expected sensitivity (as in Note 1).
(6) White noise power derived from standard deviation of single frames.
(7) White noise power derived from standard deviation of sky map.
(8) White noise power derived from fit to power spectrum.
}
\end{deluxetable}
}
\twocolumngrid

\section{Summary} \label{sec:summary}

We have performed an experiment specifically designed to 
measure the impact of the zodiacal light on the estimate
of the spatial fluctuations of the CIB. To provide the 
greatest possible sensitivity to zodiacal light effect, 
our test monitored a fixed patch in the COSMOS field
at low ecliptic \edit1{latitude} as the mean zodiacal light intensity varied
over the full accessible range of brightness (or solar elongation).
The CIB spatial power spectrum was calculated at 5 epochs
over this 5-week interval. The power spectra are characterized 
as the sum of (a) a white noise component, (b) a sky shot
noise component, and (c) a power \edit1{law} component dominating
on large angular scales.

We find that approximately half of the white noise component 
is correlated with the \edit1{varying} mean intensity of the zodiacal light.
Photon shot noise of the zodiacal light is expected to be 
the main contribution to this correlated component. 
\edit1{Detailed analysis of the non-zodiacal light portion of 
the white noise is limited by}
inaccuracies of the zodiacal light model in predicting the intensity 
in this direction as seen from {\it Spitzer}'s location within 
the interplanetary dust cloud.

The sky shot noise in the angular power spectra of the background 
is not reliably distinguished in the 
relatively shallow observations of this experiment.
The power law component does not show significant correlation 
with the mean zodiacal light intensity at 3.6 or 4.5 $\mu$m.
This confirms that observed spatial fluctuations at large scales 
($\gtrsim100''$) are not being influenced by zodiacal light.
Prior observations had been less conclusive because they 
were usually limited to high ecliptic latitudes, where the 
zodiacal light is faintest, and to epochs $\sim6$ or 12 months 
apart, where there should be minimal modulation of the zodiacal
light intensity.

\acknowledgments
\edit1{We thank the referee, M. Zemcov, for comments that improved this paper.
Additional helpful comments were provided by M. Ashby.}
This work is based on observations made with the {\it Spitzer Space Telescope}, 
which is operated by the Jet Propulsion Laboratory, California Institute of Technology 
under a contract with NASA. 
This work was funded by JPL under {\it Spitzer} Cycle 8 funding contract 
1464716, and in part through NASA/12-EUCLID11-0003 ``LIBRAE: Looking at 
Infrared Background Radiation Anisotropies with Euclid''.
This research has made use of NASA's Astrophysics Data System Bibliographic 
Services, and the IDL Astronomy Library \citep{landsman:1993}.

\facility{Spitzer (IRAC)} 
\software{IDL} 

\bibliographystyle{aasjournal}
\bibliography{zodi-cib}

\appendix
\section{Temporal Variability of the IRAC Data}

Figure \ref{fig:Fq} showed that the temporally variable offset, $F^q$, 
derived by the self-calibration is accounting for variations apart
from the simple trends expected of the zodiacal light. In this appendix,
we demonstrate additional real trends that are being found and subtracted 
by the $F^q$ term in the self-calibration.

\subsection{Zodiacal Light}
The main trend expected in $F^q$ is the temporal variation of the 
zodiacal light as the solar elongation of the target field steadily increases.
We model this trend as a constant times the model zodiacal light brightness
(as specified by the ZODY\_EST keywords in the BCD headers).

\subsection{First-Frame Effect}
Apart from the zodiacal light the strongest trend in $F^q$ is due to 
the ``first frame effect'' in the IRAC 
detectors\footnote{\url{http://irsa.ipac.caltech.edu/data/SPITZER/docs/irac/iracinstrumenthandbook/}}. 
This effect
appears as a variation in the background level as a function of the time
since the previous exposure. It appears most strongly at the first frame 
in an AOR, when there has been a long slew from the previous target. 
The BCD pipeline attempts to correct for the first frame effect, but 
is not entirely successful. Here we model the first frame effect as a 
third-order polynomial function of the delay since the preceding frame
(as specified by the FRAMEDLY keywords in the BCD headers).

\subsection{Spatial Gradient}
The zodiacal light has an intrinsic spatial gradient, being brighter at 
smaller solar elongations. During the course of dithering and moving between 
each of the $2\times 2$ fields of view at each epoch, locations at slightly
higher and lower elongations are sampled. The corresponding variations in 
brightness are thus mapped as temporal variations in $F^q$. Additionally,
the self-calibration is degenerate with respect to linear gradients across the 
field, which would have the same effect as the intrinsic zodiacal light 
gradient. We model spatial gradients in $F^q$ as a constant plus linear 
function of the $x$ and $y$ coordinates. The choice of the coordinate system is
irrelevant, as the coefficients will simply adjust appropriately for any 
chosen system.

\subsection{Exponential Decay}
The final systematic effect evident in $F^q$ is an apparent decaying 
response (with a negative amplitude) across each AOR. This trend can be 
fit by a simple exponential 
decay as a function of frame number in each AOR. However, there are both  
fast and slow decay terms with $e$-folding constants of 9 and 70 frames. 
We model this decay as a linear combination of these two 
exponential decays. This effect was also noted by \cite{krick:2011}, but it
appears more cleanly here.

The net model for the temporal variation in $F^q$ is thus given by:

\begin{eqnarray} \label{eq:offsets}
F^q_{\rm model} & = & A\ ZODY\_EST +\\ \nonumber
      &   & B_1\ FRAMEDLY + B_2\ FRAMEDLY^2 + B_3\ FRAMEDLY^3 +\\ \nonumber
      &   & C + D_x\ x + D_y\ y +\\ \nonumber
      &   & E_9\ e^{-frame/9.} + E_{70}\ e^{-frame/70.}
\end{eqnarray}

\noindent

\begin{figure}[b] 
   \centering
   \includegraphics[width=3.3in]{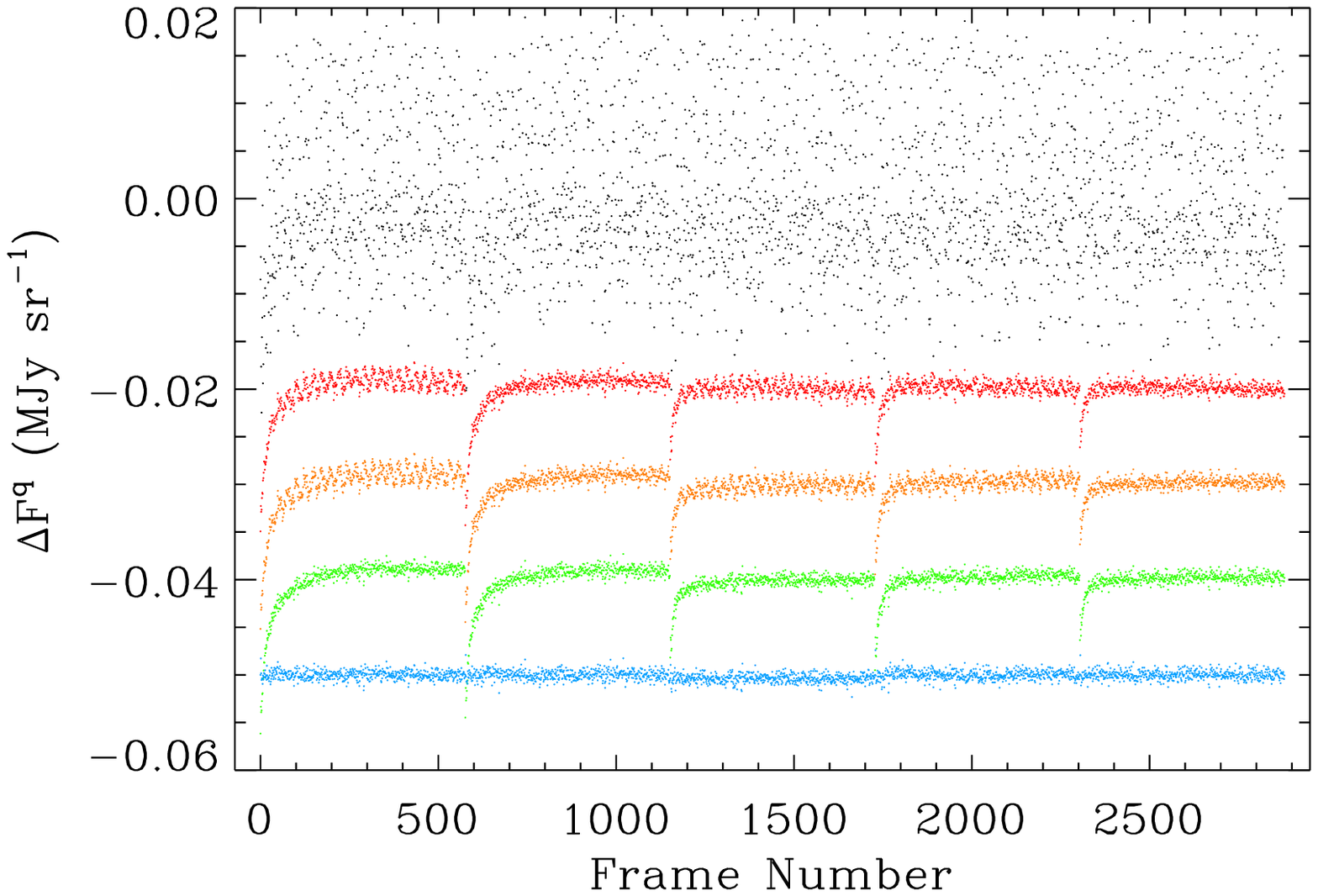}\hspace{0.4in}
   \includegraphics[width=3.3in]{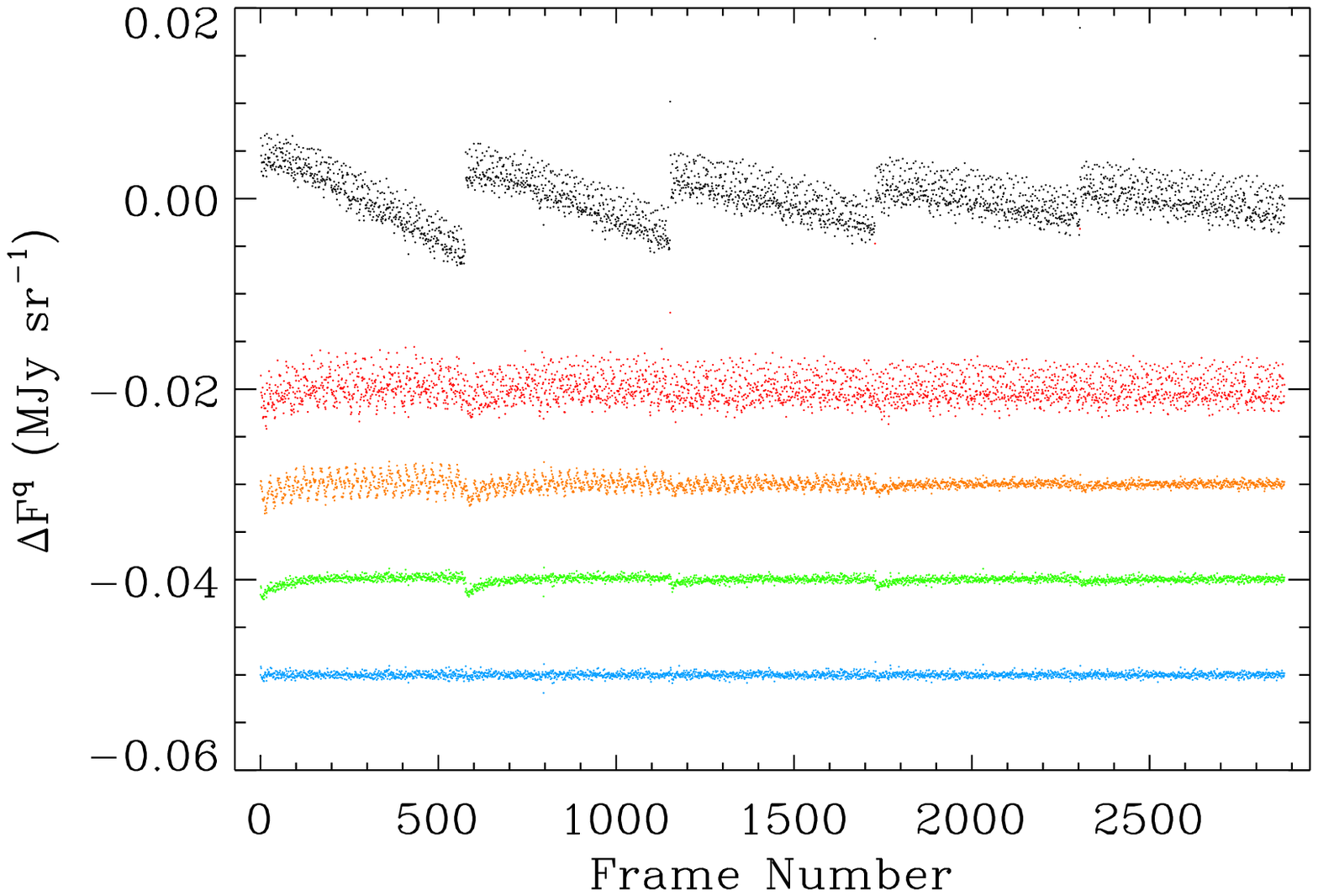} 
   \caption{The successive removal of identifiable components of
   $F^q$. For 3.6 $\mu$m (left) the black dots show the full $F^q$
   (as in Figure \ref{fig:Fq}). 
   Each segment of 576 frames corresponds to one AOR.
   The colored dots show the residual $F^q$
   after successive subtraction decreasingly smaller components: 
   a first-frame effect (red), 
   the zodiacal light variation (orange),
   an arbitrary linear gradient (green),
   and, an exponential recovery term (blue). 
   At 4.9 $\mu$m (right), the relative strength of the components differs 
   and sequence shown is after successive subtractions of:  
   the zodiacal light variation (red),
   a first-frame effect (orange), 
   an arbitrary linear gradient (green),
   and an exponential recovery term (blue).
   \label{fig:Fq-components}}
\end{figure}

Figure \ref{fig:Fq-components} shows the residual $F^q$ after successive 
subtraction of each of the components of $F^q_{\it model}$ (with arbitrary 
offsets for clarity). 

At 3.6 $\mu$m the first-frame effect is responsible for most of the 
variance in $F^q$, as seen by comparing the derived $F^q$ (black dots) 
with the derived $F^q$ minus the fitted first frame effect (red dots). 
At 4.5 $\mu$m, the zodiacal light trend is clearly dominant. Therefore
at 4.5 $\mu$m the red dots represent the derived $F^q$ minus the fitted 
zodiacal light trend. 

The comparison of the red and orange dots in both 
panel shows the subsequent subtraction of the zodiacal light trend 
(3.6 $\mu$m) and the first frame effect (4.5 $\mu$m). The zodiacal light
trend has little effect at 3.6 $\mu$m. The influence of the 
first frame effect is now
clear at 4.5 $\mu$m, though far smaller in amplitude than at 3.6 $\mu$m.

Comparison between the orange and green dots shows the subsequent subtraction
of the spatial gradient terms at both wavelengths. The AORs were designed 
such that the spatial gradients map into oscillations with a period of 
24 frames. This makes them distinguishable from the slower monotonic change
in the zodiacal light intensity which occurs as a function of time.
The amplitudes of the gradient terms are visibly larger in the earliest
AOR and decrease across AORs as does the zodiacal light intensity.

The green dots clearly show the exponential decay behavior. At 3.6 $\mu$m,
the slow 70-frame decay is evident in the first AORs, but there is a 
gradual transition to the faster 9-frame decay in the later AORs. 
At 4.5 $\mu$m, the amplitude of the decay is much smaller, and the 
slow 70-frame decay is dominant for all AORs.

After removal of the exponential decays, the blue dots show a 
fairly random distribution, with a standard deviation that is $>10$ times
smaller than present in the original $F^q$. These residual dispersions of 
$\sigma = 4.7\times 10^{-4}$ and $2.5\times 10^{-4}$ MJy sr$^{-1}$
represent an upper limit on the accuracy of the $F^q$ term of the 
self-calibration. The residual variation may still be accounting for real
effects in the data, but the specific nature of such effects has not been 
identified here.

\section{Model Depth}

\begin{figure*}[p] 
   \begin{center}
   \includegraphics[height=8.5in]{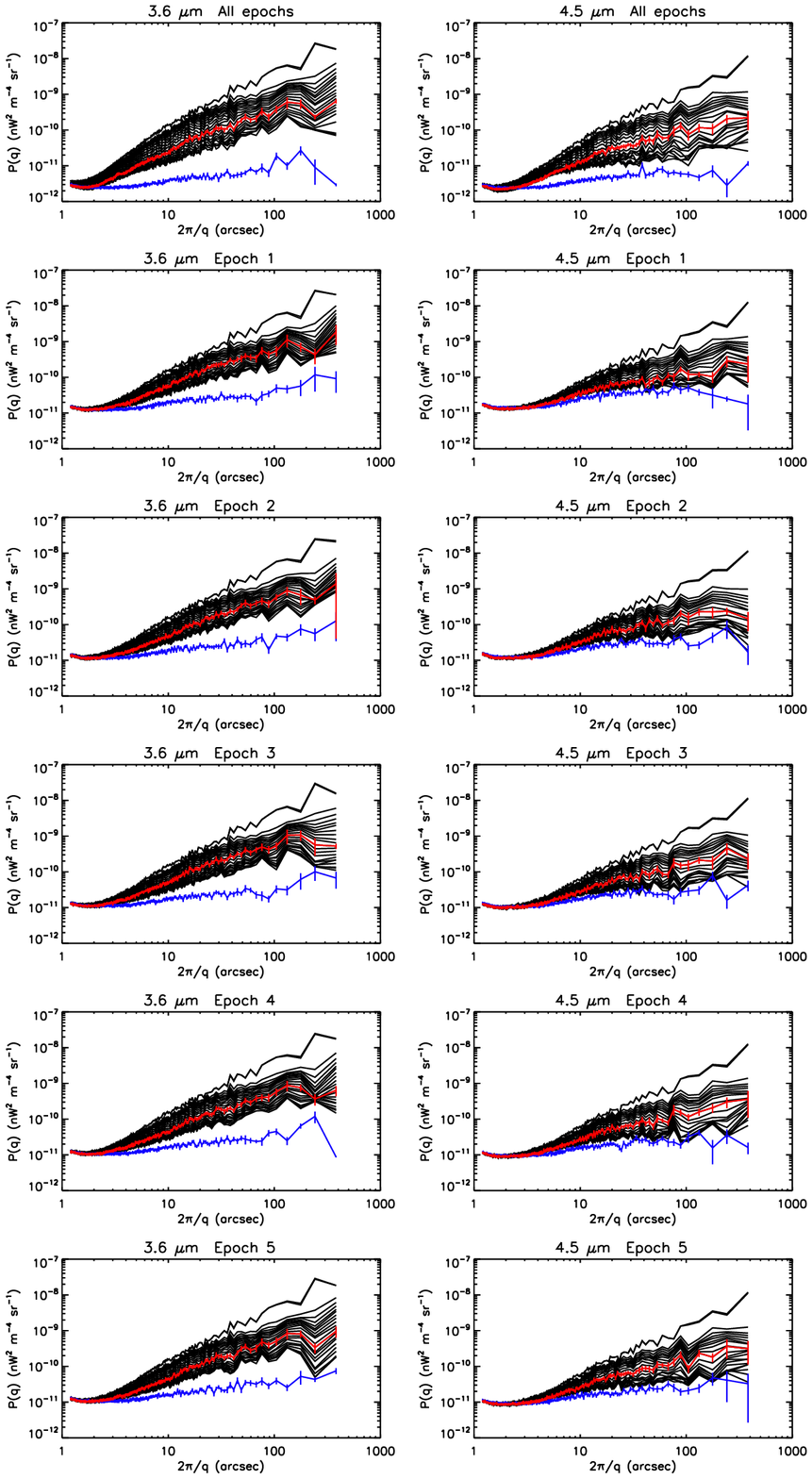}
   \end{center}
   \caption{Power spectra shown at various model depths for the combined
   and separate epochs, at 3.6 $\mu$m (left) and 4.5 $\mu$m (right).
   The black lines show that the large scale power does drop as \edit1{the}
   source model is pushed deeper. The red lines show the power spectra
   at ``optimal'' model depth when the skewness of the background (excluding 
   outliers) is zero. The blue lines show the $A-B$ power spectra.
   \label{fig:model_depth}}
\end{figure*}

A critical aspect of this analysis is the use of a source model to
remove the effects of (a) the emission of extended sources and
PSF wings that project beyond the masked areas, and (b) faint sources
which cannot be masked without adversely decreasing the fraction
of area available for analysis.
Figure \ref{fig:model_depth} shows power spectra of the masked 
images as the depth of the source model (i.e. the number of 
components subtracted) is linearly increased. 
Our choice is to use the model depth at which the skewness of the 
intensities of the unmasked pixel is zero. This assumes that a
positive skewness signifies the presence of residual sources
in the image, and that a negative skewness indicates that the 
model has begun subtracting the positive side of the noise 
distribution rather than actual sources.
The zero-skewness models are evaluated independently for each epoch, and 
are indicated by red lines (with error bars)
in Figure \ref{fig:model_depth}. 
The power spectra of the $A-B$ images are indicated by the blue lines.
This figure is analogous to Figures 8 -- 13 of \cite{arendt:2010a}.

\section{Comparison to Prior Results}

\begin{figure}[t] 
   \begin{center}
   \includegraphics[width=5in]{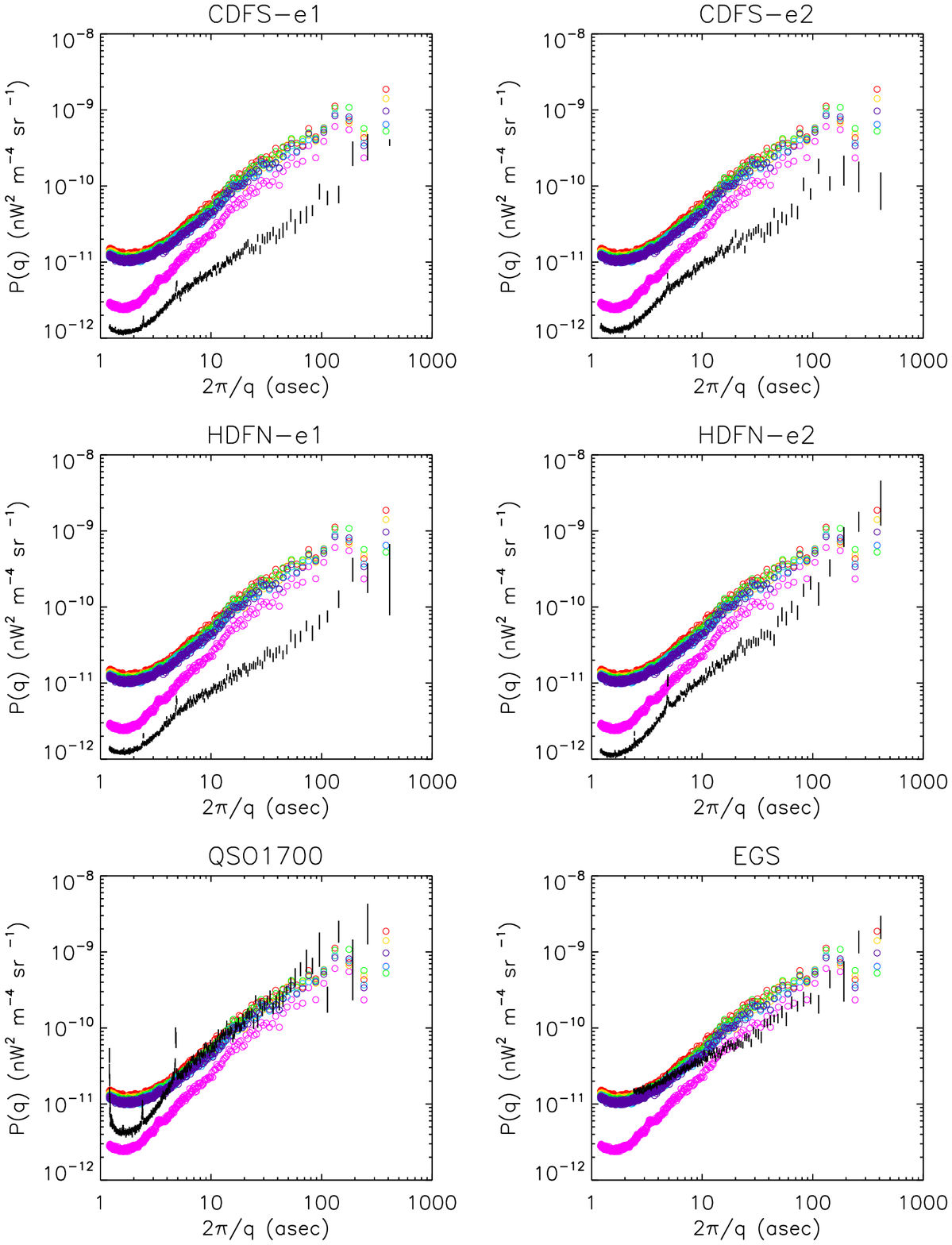}
   \end{center}
   \caption{Comparison of the 3.6 $\mu$m power spectra measured here (red, orange, green, 
   blue, violet, and magenta = Epochs 1, 2, 3, 4, 5, and all epochs combined),
   with the power spectra measured in 6 fields from \cite{arendt:2010a} as indicated
   by the black error bars.
   \label{fig:compare1}}
\end{figure}

Figure \ref{fig:compare1} compares the power spectra measured here with those
measured previously in \cite{arendt:2010a}. The previous power spectra are 
indicated by the black error bars. The power spectra measured here for 
epochs 1 -- 5 are rainbow colored from red to blue. Magenta symbols indicate 
all epochs combined. The smallest scale power (dominated by white noise) 
decreases appropriately when all epochs are combined, whereas the large scale power
is not strongly reduced by combining epochs. The large scale power 
is much higher than that seen in the deeper CDFS and HDFN observations, 
but is comparable to that of the shallower QSO1700 and EGS fields.

\begin{figure}[t] 
   \begin{center}
   \includegraphics[width=5in]{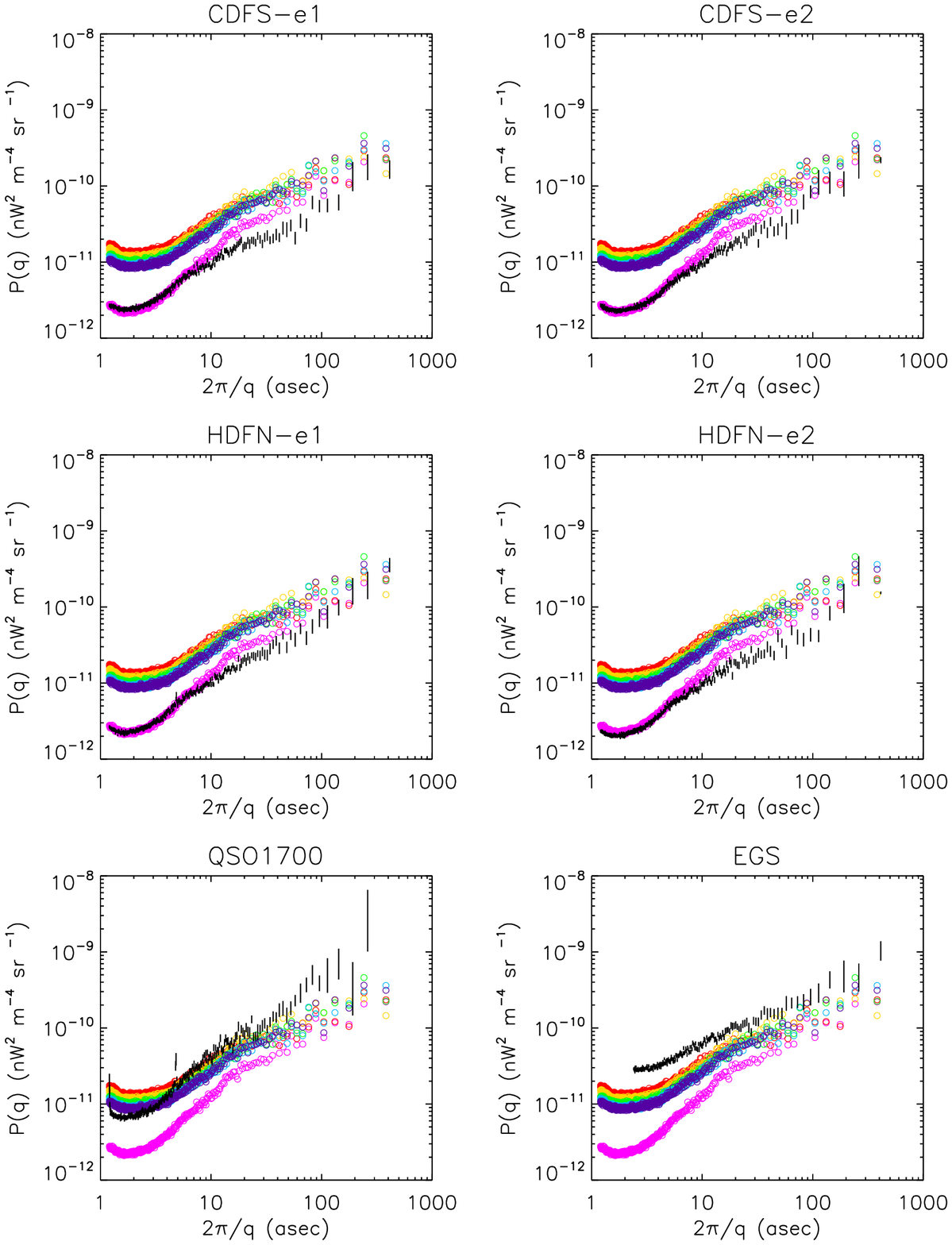}
   \end{center}
   \caption{As Figure \ref{fig:compare1}, but for 4.5 $\mu$m power spectra.
   \label{fig:compare2}}
\end{figure}

Figure \ref{fig:compare2} shows the same comparisons at 4.5 $\mu$m.
In this case, the shot noise level of the combined epochs is similar to 
that observed in the deep CDFS and HDFN fields. The large scale power 
is also close to the levels measured in these deep fields.

\end{document}